\documentclass[12pt]{article}
\usepackage{amsmath}

\newcommand*{\Pm}{\mbox{Pm}}
\newcommand*{\LL}{\mbox{L}}
\newcommand*{\TT}{\mbox{T}}
\newcommand*{\E}{\mbox{E}}
\newcommand*{\X}{\mbox{X}}
\newcommand*{\F}{\mbox{F}}
\newcommand*{\dd}{\mbox{d}}
\usepackage{natbib}
\usepackage{graphicx}

\title{What can Hurricane SAM (2021) tell us about ocean waves under tropical cyclones?}

\author{X. Zhao$^{1,2}$, L. Oruba$^{2}$, D. Hauser$^{2}$, B. Zhang$^{1}$\ and E. Dormy$^{3}$}

\date{\hspace{0.3cm}\footnotesize $^1${School of Marine Sciences, Nanjing University of Information Science and Technology, Nanjing, China}\\
$^2${Laboratoire Atmosphères, Observations Spatiales (LATMOS), Sorbonne Université, UVSQ, CNRS, Paris, France}\\
$^{3}${Département de Mathématiques et Applications, Ecole Normale Supérieure, CNRS, PSL University, Paris, France}}

\begin{document}

\maketitle

\begin{abstract}
  We investigate the ocean wave field under Hurricane SAM (2021). Whilst measurements of waves under Tropical Cyclones (TCs) are rare, an unusually large number of quality {\it in situ} and remote measurements are available in that case. First, we highlight the good consistency between the wave spectra provided by the Surface Waves Investigation and Monitoring (SWIM) instrument onboard the China-France Oceanography Satellite (CFOSAT), the {\it in situ} spectra measured by National Data Buoy Center (NDBC) buoys, and a saildrone. The impact of strong rains on SWIM spectra is then further investigated. We show that whereas the rain definitely affects the normalized radar cross section, both the innovative technology (beam rotating scanning geometry) and the post-processing processes applied to retrieve the 2D wave spectra ensure a good quality of the resulting wave spectra, even in heavy rain conditions. On this basis, the satellite, airborne and {\it in situ} observations are confronted to the analytical model proposed by Kudryavtsev et al (2015). We show that a trapped wave mechanism may be invoked to explain the large significant wave height observed in the right front quadrant of Hurricane SAM.
 \end{abstract}

\section*{Keypoints}
\begin{itemize}
\item
  We take advantage of the unusually large number of observations in
  Hurricane SAM (2021) to study the physics of waves under a
  hurricane.
\item
  The innovative CFOSAT-SWIM instrument provides 2D wave spectra which are reliable despite the heavy rain conditions in hurricanes.
\item
  The trapped wave mechanism, or extended fetch, senario can be successfully compared with observations by a saildrone and a NDBC buoy.
\end{itemize}

\section*{Plain Language Summary}

Wave measurements in Tropical Cyclones (TCs) can be performed by radars
onboard satellites or aircrafts, and by {\it in situ} devices such as buoys
or saildrones. In 2021, a saildrone sailed in Hurricane SAM, providing the
first video images of extreme weather in a hurricane, as well as wind and
wave measurements. Hurricane SAM was also monitored by the Surface Waves
Investigation and Monitoring (SWIM) instrument onboard the China-France
Oceanography Satellite (CFOSAT), which measures the wave energy
distribution. We explain why these measurements are reliable, despite
the heavy rain conditions. Using the saildrone measurements, we also show
that the large waves observed in Hurricane SAM may result from the
so-called trapped wave mechanism, whereby waves traveling at a velocity
similar to the TC displacement can undergo a phenomenal growth.

\section{Introduction}

Ocean wave measurements under Tropical Cyclones (TCs) are essential to improve our understanding of the generation of waves by TCs. They are unfortunately relatively rare and the reliability of the few existing measurements is often questioned because of the extreme wind and rain conditions inside TCs. 

    Gravity waves induced by the wind blowing at the surface of the ocean can be measured by {\it in situ} devices or by radars onboard aircrafts or satellites. The network of the National Data Buoy Center (NDBC) provides measurements of  the Significant Wave Height (SWH), their wavelength and their direction of propagation, as well as measurements of the surface wind. They are a valuable part of the US hurricane warning system and also provide precious observations for research. 
In 2021, the National Oceanic and Atmospheric Administration (NOAA) and the Saildrone US company have deployed 5 explorer uncrewed surface vehicles designed to make measurements of physical parameters, including waves, within hurricanes, between July and October 2021. On 30 September, one of the saildrones (SDs), SD 1045, crossed Hurricane SAM, on which the present paper is focused, and provided the first video of extreme conditions in a major hurricane. 
    
    Since 2019, the Chinese-French Oceanography SATellite (CFOSAT) provides
    directional wave spectra at the global scale thanks to a rotating
    multi-beam radar called SWIM (Surface Waves Investigation and
    Monitoring). These spectra cover the wavelength range $[70-500]$~m,
    giving access to both the wind-sea and the swell waves \citep{Hauser21}.
    So far, wave measurements were performed by satellite radar altimeters,
    providing a global coverage of significant wave height, and by
    Synthetic Aperture Radars (SAR) onboard satellites (e.g. Sentinel),
    providing two-dimensional spectra. These spectra are however limited by
    the so-called azimuth cut-off effect which prevents measurements of the
    wind sea waves or short swell propagating in directions close to the
    azimuth direction \citep[see][for a review]{Hauser23}.
    The CFOSAT-SWIM provides an unprecedented detailed
    description of the waves, offering a chance to measure the wind sea
    waves. A finer estimate of these waves is crucial for the investigation
    of wind-waves generated by extreme events like TCs
    \citep[e.g.][]{Oruba22}.  The CFOSAT satellite also carries a
    SCATterometer (CSCAT) providing surface wind measurements, collocated
    with the wave measurements by SWIM. Since 2008, directional wave
    spectra are also provided by the NOAA Wide Swath Radar Altimeter (WSRA)
    aboard hurricane reconnaissance aircrafts. Several reconnaissance
    flights were operated by the NOAA in the case of Hurricane SAM. 
    
    The impact of rain on satellite wind and wave products is a major issue.
    The electromagnetic waves are attenuated and scattered by the atmosphere, mainly due to the presence of water vapor and liquid water (clouds and rain), high frequencies (X-Ku-K-Ka bands) being more impacted than low frequencies (C-S-L bands).
    Moreover, raindrops impinging onto the sea surface can generate
ring waves and turbulence, thus altering the surface roughness.
    It has been shown by \cite{Quilfen06}  \citep[see also][]{Quilfen10} that reliable estimates of SWH, wind speed and rain rate can be obtained in tropical cyclones, thanks to dual-frequency altimeters, operating in both C and Ku band microwave frequencies.
    Concerning SAR observations in C-band, \cite{Alpersetal16} have shown the important and
      complex impact of rain on the normalized radar cross sections
      at medium incidence, whereas \cite{Zhao21} have analyzed more specifically the impact of rain on wave
      spectra retrieved  in tropical cyclones. The results of \cite{Zhao21} show an important impact of rain
      on the accuracy of the retrieved SWH at all the studied incidences
      (from about $20^\circ$ to about $50^\circ$). They also show that due to the rain impact, the shape of the
      wave spectra estimated from observations at medium incidence angles (around $40^\circ$)
      depart significantly from their reference spectra.

The rain impact on SWIM measurements (performed in the Ku band) has, so
far, not been investigated. 
However, SWIM products have been used in several studies on the wave field in TCs. Among them, \cite{Yurovskaya22} analysed the tropical cyclone Goni (2020) by combining multi-satellite observations (including SWIM wave spectra) and a 2D parametric wave model.  Other studies are based on a statistical approach, using data over one to several years at the global scale.  \cite{Xiang22} performed a quantitative comparison of the SWIM and CSCAT measurements in TCs to other satellite measurements and model outputs, whereas \cite{Shi21} and \cite{LeMerle22} investigated the wave distribution in TCs. \cite{Shi21} highlighted the SWH asymmetry in TCs, with the highest waves on the right of the TC track in the North Hemisphere (on the left in the South one).  The SWH's asymmetry was also shown to mainly depend on the TC intensity, with a decreasing SWH's asymmetry as the TC intensifies. Using 3 years of SWIM data in the North Hemisphere, \cite{LeMerle22} performed an analysis of the wave characteristics in $67$ TCs, by classifying them in three categories (slow, moderate and fast speed), depending on the ratio between the maximum sustained wind and the displacement velocity. The highest SWH were found in the left-front quadrant for slow-moving TCs, in the right-front quadrant for moderate speed TCs and in the right rear quadrant for fast-moving TCs.

Beyond these recent studies involving the SWIM wave spectra, the wave field
in TCs, and in particular its asymmetry, has been investigated in many
previous studies \citep[see][for a review]{Young17}. The SWH's asymmetry
partly results from the  asymmetry of the wind field.  However, an other
source of asymmetry, identified in previous studies as extended fetch or
trapped fetch \citep[e.g.][]{Young88,Bowyer05}, can also be at stake. In the
right quadrants, waves are indeed subject to high wind forcing conditions
for longer periods than usual, because of the translation of the TC:
because they move forward with the TC, they experience an extended
fetch. The translation speed of the TC plays a critical role in such a wave
containment. On the contrary, waves in the left quadrants propagate
southward, opposite to the cyclone motion: in this quadrant, the motion of
the TC tends to reduce the fetch duration. A one-dimensional analytical
model for wave field evolution forced by wind in the right and left
quadrants of a moving TC has been proposed by \cite{Kudry15}.  This
analytical model (hereafter KGC15 model) is an extension of the
self-similarity theory for wave growth \citep[e.g.][]{Badulin07} to a moving frame of reference: that 
of the TC.  In the KGC15 model, the wind blows along the direction parallel
to the TC axis with a constant speed and generates waves propagating along
the same direction. It has been recently extended to a numerical Lagrangian
model of the evolution of wave properties along wave rays propagating in TC
varying wind fields \citep{Kudry21a,Kudry21b}.  
Understanding the wave distribution in TCs is still an open issue.
In the present paper, the existence of trapped waves in hurricane SAM will
be investigated using observational data.

This paper is a multi-data sources investigation of Hurricane SAM, a
category 4 major hurricane formed on 22 September 2021 and dissipated on
$7$ October $2021$, with maximum sustained winds of $70$~m$.$s$^{-1}$. Our
study is based on satellite, airborne and {\it in situ} measurements,
including NDBC buoys and a saildrone which crossed SAM while it was in
category 4. To the best of our knowledge, this is the first time the wave
measurements in a TC by a saildrone are used in this context.   The
datasets and the methods are detailed in Section \ref{DataMethod}.  A
detailed comparison of SWIM wave spectra and {\it in situ} observations in
SAM as well as in $8$ other hurricanes and tropical storms is then
undertaken in Section  \ref{Comparison}. Section \ref{RainSWIM}  aims at
demonstrating why despite the heavy rain conditions in TCs, the SWIM wave
spectra are still reliable in TCs.   On the basis of that work, the
existence of trapped waves in SAM is put to the test in Section
\ref{Physics}. The results are discussed in Section \ref{Conclusion}.  

\section{Datasets and methods}
\label{DataMethod}

The parameters of tropical storms and hurricanes are available in the International Best Track Archive for Climate Stewardship (IBTrACS, version 4) database (Knapp et al 2010).  Best track data from the US agency,  providing 3-hourly data including locations and maximum sustained wind speed, were used here. 

\begin{figure}
 \centering
 \includegraphics[width=0.7\textwidth]{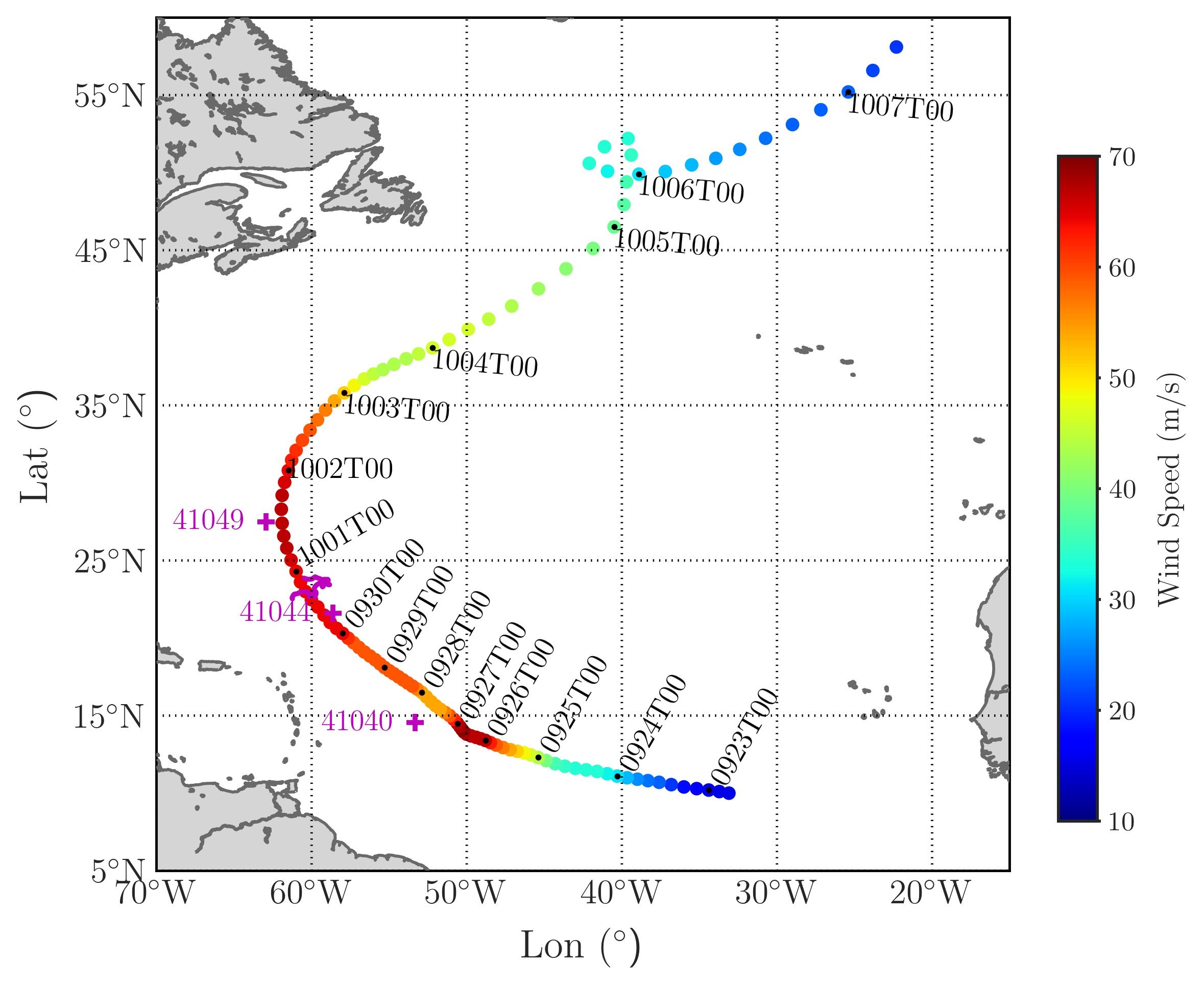}
 \caption{Trajectory of Hurricane SAM between 22 September 2021 and 7 October 2021. Colors: maximum sustained wind speed in m$\cdot$s$^{-1}$(IBTrACs data). Magenta crosses: location of the NDBC buoys 41040, 41044 and 41049. Magenta line: trajectory of the SD
   1045 during the same period.}
 \label{traj}
 \end{figure}

Figure \ref{traj} shows the trajectory of SAM between 22 September and 7 October 2021. Its maximum sustained winds were larger than $50$~m$\cdot$s$^{-1}$ (setting it in category 3) between 25 September 09:00 and $3$ October 00:00. It strengthened to category 4 (winds larger than $58$~m$\cdot$s$^{-1}$) first between 25 September 18:00 and 27 September 06:00, and then between 28 September 06:00 and 2 October 12:00.  The locations of the NDBC buoys and the trajectory of the SD 1045 between 22 September and 7 October are indicated in magenta in Figure \ref{traj}. A few hours after its first intense phase, the  track of SAM passed within $188$~km east of the NDBC buoy 41040, whereas during its second intense phase, its passed within $60$~km west of the NDBC buoy 41044, $30$~km west of the saildrone 1045 and $103$~km east of the buoy $41049$.   

The results presented in Section \ref{2Dspec} also rely on SWIM and NDBC waves measurements in four tropical storms (Sebastien in 2019, Josephine in 2020, Claudette and Fred in 2021), one hurricane of Category 1 (Isaias in 2020),  two hurricanes of Category $2$ (Sally in 2020 and Earl in 2022) and one hurricane of Category 4 (Ida in 2021).  

\subsection{{\it In situ} observations}

The NDBC buoys provide hourly one-dimensional frequency wave spectra typically from $0.03$~Hz to $0.40$~Hz as well as $5$ Fourier coefficients corresponding to the directional energy distribution. Two-directional  frequency wave spectra are reconstructed using the Maximum Entropy Estimation (MEM) method \citep{Lygre86}. Wave parameters, including the SWH, the dominant wave period and the corresponding direction of propagation, are also provided on an hourly basis. The SD data include only the SWH and the peak period of waves every $30$~minutes. Unfortunately, no information about the spectral distribution or the direction of propagation of waves is provided by the SD. NDBC buoys (resp. SD) also provide wind measurements (speed and direction) every 10 minutes (resp. 30 minutes). These measurements are performed at the anemometer height. For a proper comparison with other wind measurements,  the corresponding $10$~m height winds were retrieved  using the correction method proposed by \cite{Zieger09}, assuming a logarithmic marine boundary layer under neutral stability conditions.
Note that the winds measured by the SD $1045$ during SAM were earlier successfully compared to satellite microwave radiometer measurements and models by \cite{Ricciardulli22}. 

\subsection{Satellite and airborne wave products}

\begin{figure}
\centering
  \includegraphics[width=0.6\textwidth]{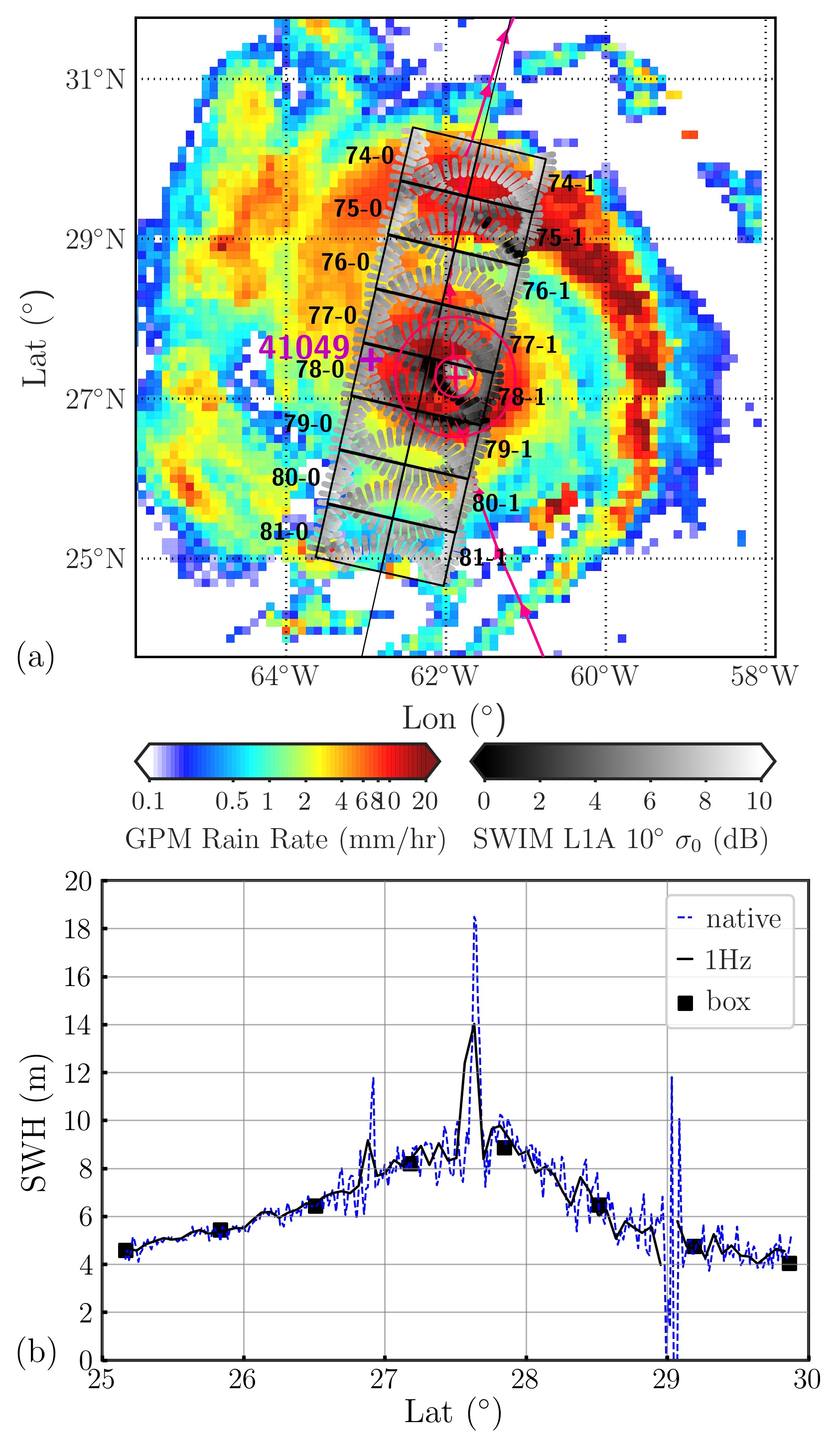}
 \caption{(a) SWIM footprints of the spectral beam at $10^\circ$ during the passage of SWIM within SAM on 1st October 11:32. Gray color: values of $\sigma_0$ (in dB) along the footprints. In black: boxes of the L2 wave products, each side of the nadir track. In colors: rain field (mm/hr) on 1st October 11:30. Red line: trajectory of the center of SAM.  Red cross: location of the center of SAM during the passage of SWIM. Red circles: $R_{max}$ and $3\, R_{max}$. Magenta cross: location of buoy 41049. (b) SWH as measured by SWIM along the nadir track:  native (dashed blue), $1$~Hz averaged (black) and averaged over each box (black squares).}
\label{Swaths_01oct}
\end{figure}

The {\it in situ} wave measurements were complemented by satellite wave
products. CFOSAT has passed several times within SAM, the passage of 1st
October 11:32 being of particular interest because it was very close to the
TC center (Figure \ref{Swaths_01oct}a). The nadir beam of SWIM provides SWH
measurements (Figure \ref{Swaths_01oct}b) whereas the $3$ off-nadir (or
spectral) beams inclined at $6^\circ$,  $8^\circ$ and $10^\circ$ with respect to nadir
provide
two-dimensional wave slope spectra. In the present study, we use the wave
spectra contained in the level 2 (L2) products
provided by the CNES Wind and Wave Instrument Center (CWWIC).  We mainly use the
products derived from the spectral beam at $10^\circ$, which tends to be
preferred by the community because it is less affected by speckle contamination
and wind \citep{Hauser21}. 

The detailed investigation of the impact of rain on SWIM wave spectra carried out in section \ref{RainSWIM} deserves some details on the geometry and post-processing chain of the SWIM instrument. First, the Normalized Radar Cross Section (NRCS) signals, hereafter denoted as $\sigma_0$, are computed within the footprint along the look direction
(about $18$~km long for the $3$ dB footprint) with a high resolution (about $8$~m when
projected on the surface). Because SWIM is a real-aperture radar, these $\sigma_0$ values
correspond to signals integrated over the azimuth direction ($3$ dB footprint of
about $18$~km). The grey segments shown in Figure \ref{Swaths_01oct}a show an example of
the sampling for successive azimuthal scans of the SWIM. Each grey segment represents
the position of the footprint in the elevation direction (the azimuth extension is not
represented here). In the post-processing, to avoid conditions of low signal to noise
ratio, only the values larger than a threshold value are kept ($3$~dB above the thermal
noise level). The $\sigma_0$ signals contain modulations due to the tilt of the long waves
when these propagate along the look direction \citep{Hauser21}. To retrieve the wave
information from the $\sigma_0$ measurements, a detrending is applied on the $\sigma_0$
profiles, and the fluctuation signals, $\delta\sigma_0$, are then calculated. 
Modulation spectra, hereafter $\Pm(k)$, are then derived from $\delta\sigma_0$ through a
Fourier transform, corrected from the speckle noise spectrum and the sensor impulse
response spectrum. The modulation spectra are then transformed into wave
slope spectra by applying a Modulation Transfer Function (MTF), which
involves a renormalization (see below). 
The different directions are finally combined to construct a two-dimensional wave slope spectrum  at the scale of a box (black rectangles in Figure \ref{Swaths_01oct}a) of about $70$~km $\times$ $90$~km, on each side of the nadir track.
As explained above, the spectral energy is normalized by using the SWH from the nadir beam. More precisely, the native nadir SWH (dashed blue curve in Figure \ref{Swaths_01oct}b) is averaged over each box, resulting in the values highlighted by the black squares in Figure \ref{Swaths_01oct}b and used to normalize the wave spectra. In this averaging process, the values larger than $3$ times the standard deviation of the series are deleted. The thus-obtained wave spectra are discretized in $32$ wave numbers in the range $[0.01–0.28]$~m$^{-1}$ and $12$~directions of propagation, with a $180^\circ$ ambiguity \citep[see][for further details]{Hauser21}. In addition to the L2 products, containing the wave spectra, section \ref{RainSWIM} also relies
on the L1A and L1B products provided by the CWWIC, which contain the $\sigma_0$, $\delta\sigma_0$ and $\Pm$ variables.
An alternative post-processing, involving running averages along the footprints, is proposed by the Ifremer Wind and Wave
Operation Center (IWWOC) to build the L2S SWIM products, containing one-dimensional wave slope spectra along the
footprints. These data are also used in section~\ref{RainSWIM}. 

Sentinel 3 has passed twice over SAM: on 30 September 01:34 and on 1st October 01:49. Hence, the present study also involves the $1$~Hz averaged L2P SWH and surface wind measurements by the Sar Radar ALtimeter (SRAL) onboard Sentinel 3.  Besides, Hurricane SAM was also sampled by the SAR of Sentinel 1. Among the passages of Sentinel 1, the passages on 2 October 09:54 and 3 October 09:45 have retained our interest, because they are very close (both in space and time) to SWIM. The level 2  products of the Sentinel 1a and 1b SAR wave mode were used. They correspond to two-dimensional wave height spectra, built from imagettes of size $20$~km $\times $ $20$~km, and discretized in $60$ wavenumber bins in the range $[0.01, 0.21]$~m$^{-1}$ and $72$ directions from $0^\circ$ to $360^\circ$.

Finally, during Hurricane SAM, $4$ reconnaissance flights were performed by NOAA. The airborne Wide Swath Radar Altimeter (WSRA), which is a radar altimeter operating at 16 GHz, provides real-time continuous wave measurements. We used the level 4 (L4) product which contains two-dimensional wave spectra $\E(k_x,k_y)$ of $65 \times 65$ spectral values, where $k_x$ and $k_y$ are the east and north components of the wavenumber vector.

\subsection{Wind and rain products}

The wave distribution in TCs is partly correlated to the surface wind
field. In addition to the wind measurements by the NDBC buoys and the SD,
our study also relies on satellite and airborne wind measurements, as well
as model outputs. We used  the CSCAT L2B wind products of 25-km grid
resolution provided by the National Satellite Ocean Application Service
(NSOAS). It was recently shown by \cite{Zhao22} that for high winds
  (above $15$ m/s) and for rain rates up to $6$~mm/hr, there is a good agreement between
  the CSCAT winds and the European Centre for Medium-Range Weather Forecasts (ECMWF) winds.
  However, it must be kept in mind that they also show that the CSCAT winds retrieved within  Typhoon Maysak (2020) are underestimated (compared to ECMWF) for heavy rain conditions
  (rain rate larger than $6$~mm/hr). 
These data are completed by
the surface wind outputs of the ECMWF Integrated
Forecasting System (IFS, version 47r1) on a $0.1^\circ \times 0.1^\circ$ grid available every $6$ hours. That is worth noting that thanks to a new roughness length parameterization (considering the roughnesslength decrease under extreme wind conditions), the ECMWF IFS numerical simulations in version 47r1 perform much better than the previous versions in capturing the high winds in tropical cyclones \citep[e.g.][]{Bidlotetal20,Lietal21}.

The rain rate (hereafter $RR$) is estimated using the Integrated Multi-satellitE Retrievals for GPM (IMERG) Final Run product. The IMERG product integrates all microwave precipitation estimates from GPM constellation and Infrared-based observations, with Global Precipitation Climatology Centre monthly gauge calibration, including a backward and forward morphing technique  \citep{Huffman23}.
The IMERG product in version 7 is half-hourly computed on a $0.1^\circ
\times 0.1^\circ$ spatial grid.
Only data with a quality index larger than $0.6$ are considered in the present study.

\subsection{Methods}
\label{Meth}

The natural spectral variable for {\it in situ} observations being the frequency, the SWIM wave slope spectra $\F(k,\theta)$ (in m$^2$) are converted to wave height spectra $\E(k,\theta)$ (in m$^4$) and then to frequency wave spectra $\widetilde{\E}(f,\theta)$ (in m$^2\cdot$s)  using the formula
\begin{equation}
    \E(k,\theta)=\frac{\F(k,\theta)}{k^2}\,, \qquad  \widetilde{\E}(f,\theta) = \E(k,\theta) k \frac{\dd k}{\dd f} \,, 
\end{equation}
and the gravity waves dispersion relation.  The SWH is estimated as 
\begin{equation}
\mbox{SWH}= 4 \sqrt{\E_{tot}} \qquad{\rm where} \qquad \E_{tot}= \iint_{0}^{\pi} \widetilde{\E}(f,\theta) \, \dd f \dd \theta \,,
\label{SWH}
\end{equation}
the double integration being performed over the directions $\theta$ and the frequencies $f$.  
The dominant frequency, $f_p$ (resp. dominant wavelength $\lambda_p$), and the dominant direction, $\theta_p$, provided in the NDBC (resp. SWIM) products have been estimated using different procedures, which prevents from proper comparisons. These parameters are thus recalculated here using the frequency wave spectra $\widetilde{\E}(f,\theta)$.
Following \cite{LeMerle22}, a two-dimensional Gaussian filter is applied to the directional spectra to suppress spurious energy peaks that can form at low wavenumbers in the SWIM spectra, before applying the following formula
\begin{equation}
f_p=\frac{ \int_{f_{max}- \Delta f}^{f_{max} + \Delta f} f \,
  \widetilde{\E}(f,\theta_{max}) \dd f}{ \int_{f_{max}- \Delta f}^{f_{max}
    + \Delta f} \widetilde{\E}(f,\theta_{max}) \dd f} \, ,
\qquad
\theta_p = \frac{ \int_{\theta_{max}- \Delta \theta}^{\theta_{max} + \Delta \theta} \theta \, \widetilde{\E}(f,\theta_{max}) \dd \theta}{ \int_{\theta_{max}- \Delta \theta}^{\theta_{max} + \Delta \theta} \widetilde{\E}(f,\theta_{max}) \dd \theta}  \,,
\end{equation}
where ($f_{max}$, $\theta_{max}$) is the location of the maximum energy,
and ($\Delta f$ , $\Delta \theta$) are the frequency and direction
intervals.
The same procedure was applied to the SAR and WSRA wave height spectra.
The direction convention for $\theta$ is the direction from which waves travel, measured clockwise from North. The dominant wavelengh $\lambda_p$ is deduced from $f_p$ using the dispersion relation.

The following study involves distance calculations between SWIM and {\it in situ} measurements. 
The fact that the SWH measurements are performed along the nadir track (black line in Figure \ref{Swaths_01oct}), whereas the wave spectra (hence the $\lambda_p$ and $\theta_p$ parameters) are computed over boxes (black quadrangles in Figure \ref{Swaths_01oct}), is taken into account when calculating the distance between measurements.

\section{Waves observations by SWIM and {\it in situ} devices}
\label{compaSWIMinsitu}

In this section, the {\it in situ} observations by NDBC buoys and the SD 1045 are used to assess the reliability of the SWIM wave spectra under heavy rain conditions in SAM. In addition, SWIM wave spectra are compared to NDBC wave spectra within $8$ other tropical storms and hurricanes. 

\subsection{Multi-sources composite description of wind and waves conditions in SAM}

\begin{figure}
 \centerline{
   \includegraphics[width=1\textwidth]{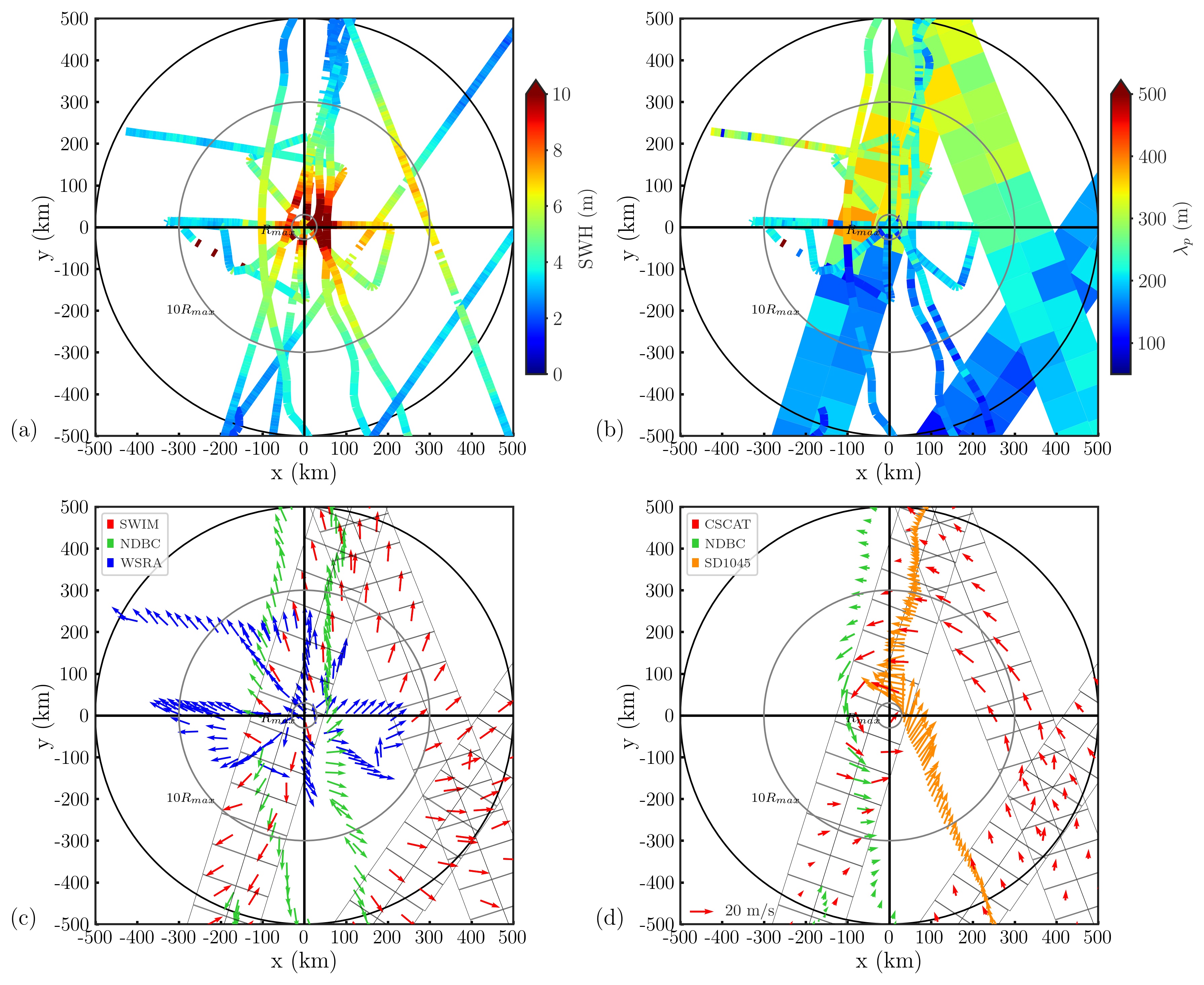}
 }
 \caption{Composite representation of observations between 29 September 00:00 and 2 October 12:00 reported in the frame of Hurricane SAM.
   (a) SWH from SWIM, Sentinel 3, the airborne WSRA radar, $3$ NDBC buoys and the SD, (b) Dominant wavelength $\lambda_p$ from SWIM (large squares), the airborne WSRA radar, $3$ NDBC buoys and the SD, (c) Dominant wave direction from SWIM, the airborne WSRA radar and $3$ NDBC buoys (see inset for color code), (d) Wind vectors from  CSCAT, $2$ NDBC buoys and the SD (see inset for color code). Black circles indicate  $R_{max}$, $10 \, R_{max}$ and a radius of $500$~km, with $R_{max}=28$~km.}
 \label{frame_SAM}
 \end{figure}

A composite over $3.5$ days is first built, using all of the available
observations, in order to characterize the wave distribution in Hurricane
SAM. Figure \ref{frame_SAM} is a composite constructed using different data
sources over $3.5$ days, between 29 September 00:00 and 2 October
12:00. That period corresponds to the second high-intensity phase of
SAM. During that period, its maximum sustained winds were comprised between
$59$~m$\cdot$s$^{-1}$ and $67$~m$\cdot$s$^{-1}$, which corresponds to a
relative change in intensity of about $10\%$.
The measurements are all reported in the frame of SAM, with the top of the figure
corresponding to the TC propagation direction.  The figure presents both
waves and winds {\it in situ} measurements by the 3 NDBC buoys ($41040$,
$41044$ and $41049$) and by the saildrone SD 1045.  The wave measurements by SWIM
during 3 passages 
(on 29 September 22:32, 1st October 11:32 and 2 October 11:16) are
superimposed, as well as the SWH as measured by the Sentinel-3 altimeter
during the two passages on 30 September 01:34 and 1st October
01:49. The WSRA wave data collected during the reconnaissance flight
performed between 29 September 20:38 and 30 September 03:03 are also
reported on the same figure.
Finally, the CSCAT winds are also plotted in
Figure \ref{frame_SAM}d, at the position of the SWIM wave boxes.

Figures \ref{frame_SAM}a, \ref{frame_SAM}b and \ref{frame_SAM}c show the
SWH, the dominant wavelength and the dominant wave directions,
respectively, whereas Figure \ref{frame_SAM}d shows the wind vectors. In
Figure  \ref{frame_SAM}c, the $180^\circ$ ambiguity inherent to the SWIM
wave spectra is raised using the CSCAT winds, by assuming that under TCs
conditions, the angle between the direction of the wind and the direction
of wave propagation is always lower than $180^\circ$, which has been
observed in many previous studies.  The largest waves are observed near the
TC center and in the right front quadrant at distances smaller than $3\,
R_{max}$ (Figure \ref{frame_SAM}a), whereas the longest wavelengths are
observed in the left front and right front quadrants (Figure
\ref{frame_SAM}b).  Waves in the two front quadrants radiate out of a
region to the right of the hurricane center, corresponding to a region of
strong wind (see Figures \ref{frame_SAM}d and \ref{Track_01oct}b
which shows the CSCAT wind field on 1st October 11:32).
It is worth noting that wind and dominant waves are never aligned.
The angle between winds and waves is about $60^\circ$ in the right front quadrant,
and larger than $60^\circ$ in most locations in the other quadrants (compare Figures \ref{frame_SAM}c and \ref{frame_SAM}d).

It is interesting to compare these results to \cite{LeMerle22}. Between 29 September 00:00 and 2 October 12:00, the displacement velocity $V$ of SAM increased from $3.6$~m$\cdot$s$^{-1}$
to $9.3$~m$\cdot$s$^{-1}$ on 1st October 18:00 before decreasing to $7.7$~m$\cdot$s$^{-1}$.
 The ratio between the maximum sustained wind $U_{max}$ and the displacement velocity $V$ was mainly comprised between $5$ and $12$, which corresponds to a moderate speed TC, according to their classification. The above description is consistent with the results obtained by \cite{LeMerle22} for moderate speed TCs.  

\subsection{SWIM and {\it in situ} estimations of wave parameters in SAM}
\label{Comparison}

In order to assess the reliability of SWIM wave measurements in Hurricane
SAM, a comparison of the wave parameters as measured by SWIM and by the
{\it in situ} devices is performed.  The data provided by the NDBC buoys
and  by the SD are considered here as the most reliable wave measurements
in SAM. The main issue raised when comparing satellite to {\it in situ}
measurements is due to the sparse nature of these two types of data,  both
in space and time. On the one side, the {\it in situ} observations are
performed at the sole location of the device, which is fixed for the NDBC
buoys and which changes over time for the SD, according to its
trajectory. These data are provided every 10 minutes for the NDBC buoys
(hourly for the spectra) and every 30 minutes for the SD. On the other
side, SWIM passages in a given region of the globe take place once or twice
a day, and the wave measurements are performed only along the nadir track
(for the SWH) and in two bands of about $90$~km wide on each side of the
track (for the wave spectra). There is therefore inevitably a temporal and
spatial difference between the SWIM and {\it in situ} measurements. Such
differences can be critical under TCs because of the inhomogeneous
distribution of waves in TCs and because of their temporal variability, due
to the wind temporal variations and to the TC displacement. Such
comparisons must thus be performed carefully, especially when assessing the
quality of satellite measurements.  

\begin{figure}
 \centering
 \includegraphics[width=1\textwidth]{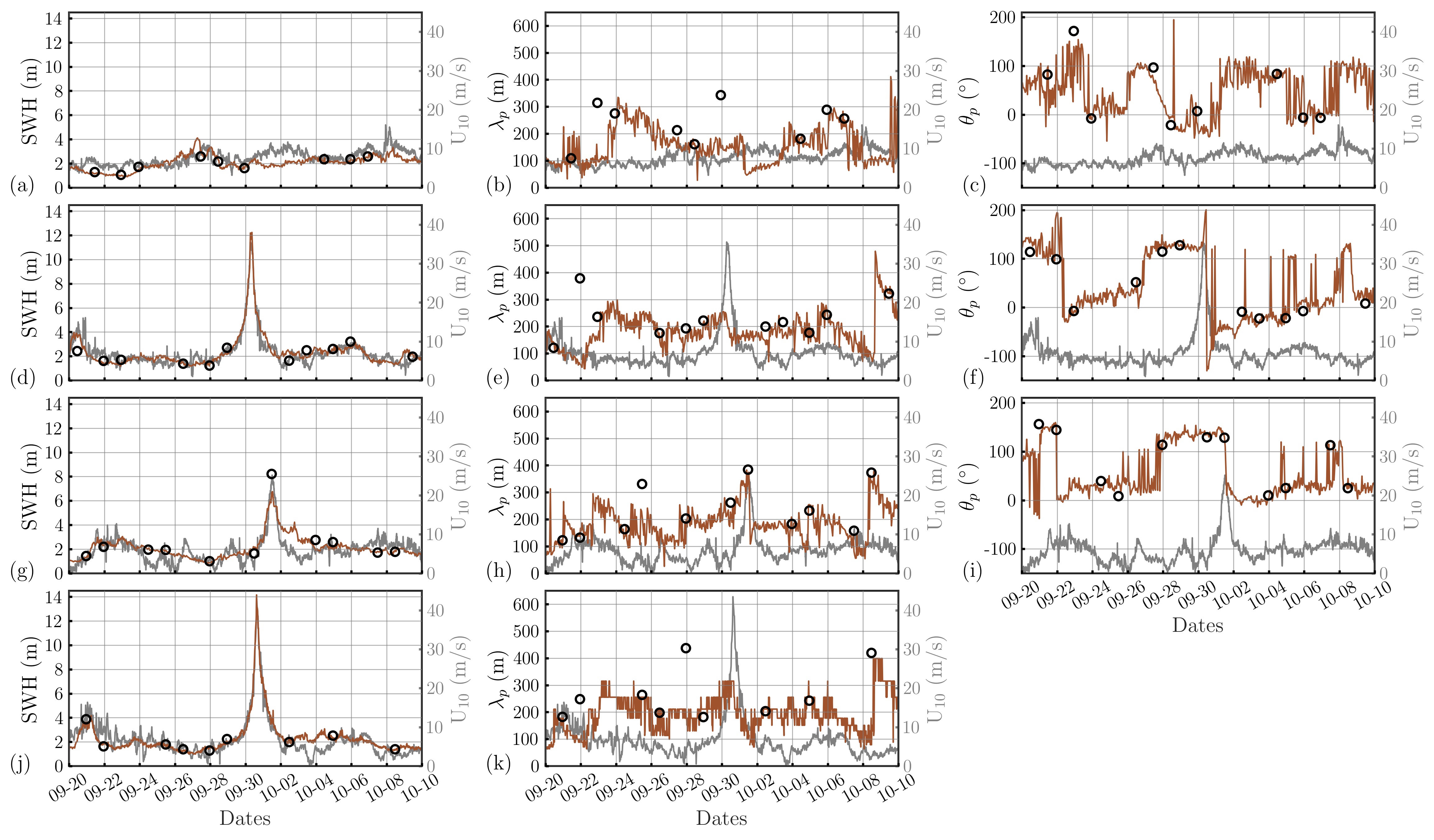}
 \caption{Brown curves: SWH (panels a,d,g,j), dominant wavelength $\lambda_p$ (panels b,e,h,k) and dominant direction $\theta_p$ (panels c,f,i) as measured by the buoys 41040 (panels a,b,c), 41044 (panels d,e,f), 41049 (panels g,h,i) and the SD 1045 (panels j,k) between 20 September and 10 October. Gray curves: wind speed. Black circles: closest SWIM measurements.}
 \label{compa_all}
 \end{figure}

The brown lines in Figure~\ref{compa_all} show the time evolution of the SWH (panels a,d,g,j), the dominant wavelength $\lambda_p$ (panels b,e,h,k) and the dominant direction $\theta_p$ (panels c,f,i) as measured by the buoys 41040 (panels a,b,c), 41044 (panels d,e,f), 41049 (panels g,h,i) and the SD 1045 (panels j,k) between 20 September and 10 October. The parameters  $\lambda_p$ and $\theta_p$ are calculated using the method described in Section \ref{Meth}. Note that the SD data do not allow to derive $\theta_p$. The gray curves correspond to the wind speed measured by the buoys and the SD. We co-localized $40$ measurements from SWIM, by allowing a maximum distance of $300$~km between the SWIM boxes and the {\it in situ} devices, and a maximum time-lapse of $30$~minutes for the buoys and of $15$~minutes for the SD. The  wave parameters derived from the SWIM wave spectra are superimposed as black circles in Figure~\ref{compa_all}. In  Figures~\ref{compa_all}c,f,i, the $180^\circ$ ambiguity on SWIM spectra was raised by selecting the direction closest to that measured by the buoy. 

Among the {\it in situ} devices, the buoy $41040$ is the farthest one from the TC track (see Figure \ref{traj}), at $188$~km west of the TC track: during the passage of SAM, it measured a SWH of $4.1$~m on 27 September 06:40 (panel a).  The largest SWH is measured by the SD (panel j): on 30 September 15:00, while it was located at $39$~km north-east of the TC track, it measured a SWH of $14.2$~m. Going back $6.3$~hours earlier, the buoy $41044$ measured a SWH of $12.2$~m, $60$~km east of the TC track (panel d). The track of SAM goes further from the buoy $41049$, located west of the track at a distance of $103$~km (see also Figure \ref{traj}): a SWH of $6.7$~m is measured by the buoy $41049$ on 1st October 11:40 (panel g).
There is an almost simultaneous occurrence between the maximum of SWH and the maximum wind speed as measured by the buoys 41044, 41049 and by the SD resp., with maximum wind speeds of $24.4$~m$\cdot$s$^{-1}$, $34.6$~m$\cdot$s$^{-1}$ and $43.5$~m$\cdot$s$^{-1}$ resp. (see panels d, g and j).
The value of SWH corresponding to the closest measurement by SWIM during its passage on 1st October 11:32 is $8.2$~m (panel g).  Unfortunately, there was no passage of SWIM in the vinicity of the buoy $41044$ and of the SD between 28 September 22:46 and 2nd October 11:18: both SWH peaks on panels d and j are thus not sampled by SWIM.

During the passage of SWIM close to the buoys 41044, 41049 and to the SD, the dominant wavelength undergoes a sharp decrease from larger values to smaller values concomitant with the maximum of wind and SWH: in a few hours on 30 September, it decreases from $250$~m to $150$~m for the buoy 41044 (panel e) and from $300$~m to $100$~m for the SD (panel k) and on 1st October from $400$~m to $100$~m for the buoy 41049 (panel h). This result is consistent with Figure \ref{frame_SAM}b: the {\it in situ} devices, first undergo the longest waves, in the front quadrants of SAM, before undergoing shorter waves in the rear quadrants. The direction of propagation of waves also undergoes an abrupt change during the passage of SAM synchronised with the maximum wind and SWH (panels f and i): the buoys 41044 and 41049 first measure waves propagating forward and then waves propagating rather backward, which is consistent with Figure  \ref{frame_SAM}c.

The comparison of wave characteristics between SWIM measurements and {\it in situ} observations over the period
shown in Figure \ref{compa_all} exhibits overall consistency.
The discrepancy between the SWIM and
{\it in situ} SWH is of order $10\%$ and does not exceed $30\%$.
The few cases where the difference in the peak values
is significant correspond either to multi-modal systems with different wavelengths
and directions but comparable energies,
or to low SWH  cases (SWH~$\simeq 1$~m) where SWIM wave measurements are
known to be less reliable.

\subsection{Two-dimensional wave spectra in heavy rain}
\label{2Dspec}

\begin{figure}
 \centering
 \includegraphics[width=1\textwidth]{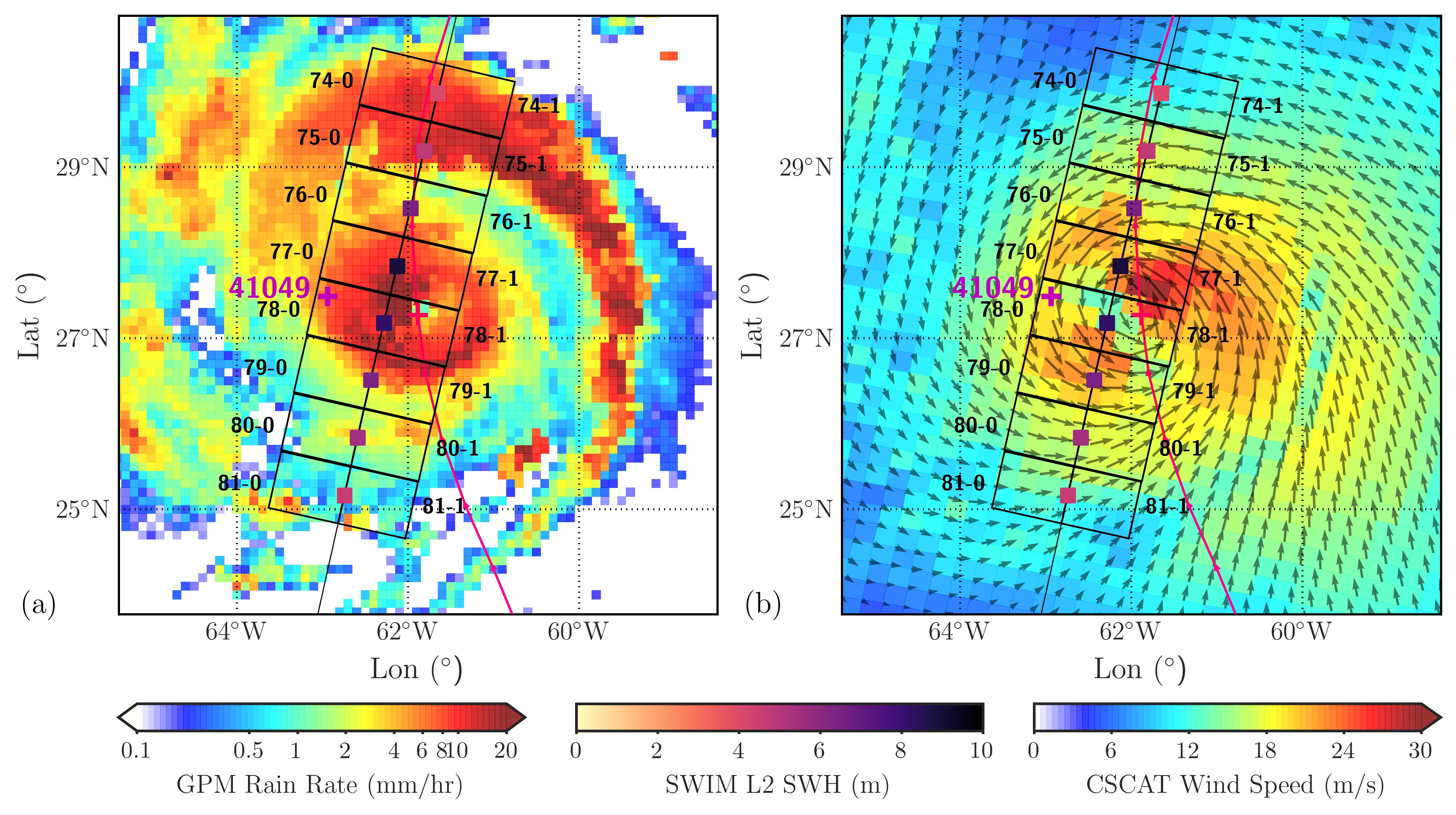} 
 \caption{SWIM nadir track (black line) and boxes (black quadrangles) during the passage of SWIM within SAM on 1st October 11:32, close to the buoy 41049 (magenta cross). 
 (a) Rain rate field (in mm/hr) on 1st October 11:30 and (b) surface wind field from CSCAT (colors: intensity in m/s, arrows: directions). Colored squares: nadir SWH averaged over boxes.
 Red line: trajectory of the center of SAM.  Red cross: location of the center of SAM during the passage of SWIM.  }
\label{Track_01oct}
 \end{figure}

During the entire period subject to investigation here, the passage of SWIM within SAM on 1st October 11:32, very close to the buoy 41049, corresponds to heavy rain along the SWIM track. In the following, we will focus on this track. Figure~\ref{Track_01oct} shows the nadir track and the boxes over which the wave spectra are computed, superimposed to the rain rate field  at 11:30 (Figure~\ref{Track_01oct}a) and to the wind field as measured by CSCAT (Figure~\ref{Track_01oct}b). The boxes numbered as 77-0, 77-1, 78-0 and 78-1 are the most severely affected by rain, with the rain rate exceeding 20 mm/hr in some areas of these boxes (Figure~\ref{Track_01oct}a). The wind speed from CSCAT does not exceed $30$~m$\cdot$s$^{-1}$, which is largely underestimated, roughly by a factor of $2$ compared to the IBTrACS estimation.
  Note that such saturation is not specific to CSCAT: it is known that at high wind speeds
  (winds larger than $30$~m$\cdot$s$^{-1}$), wind information from scatterometers saturates.

\begin{figure}
 \includegraphics[width=1\textwidth]{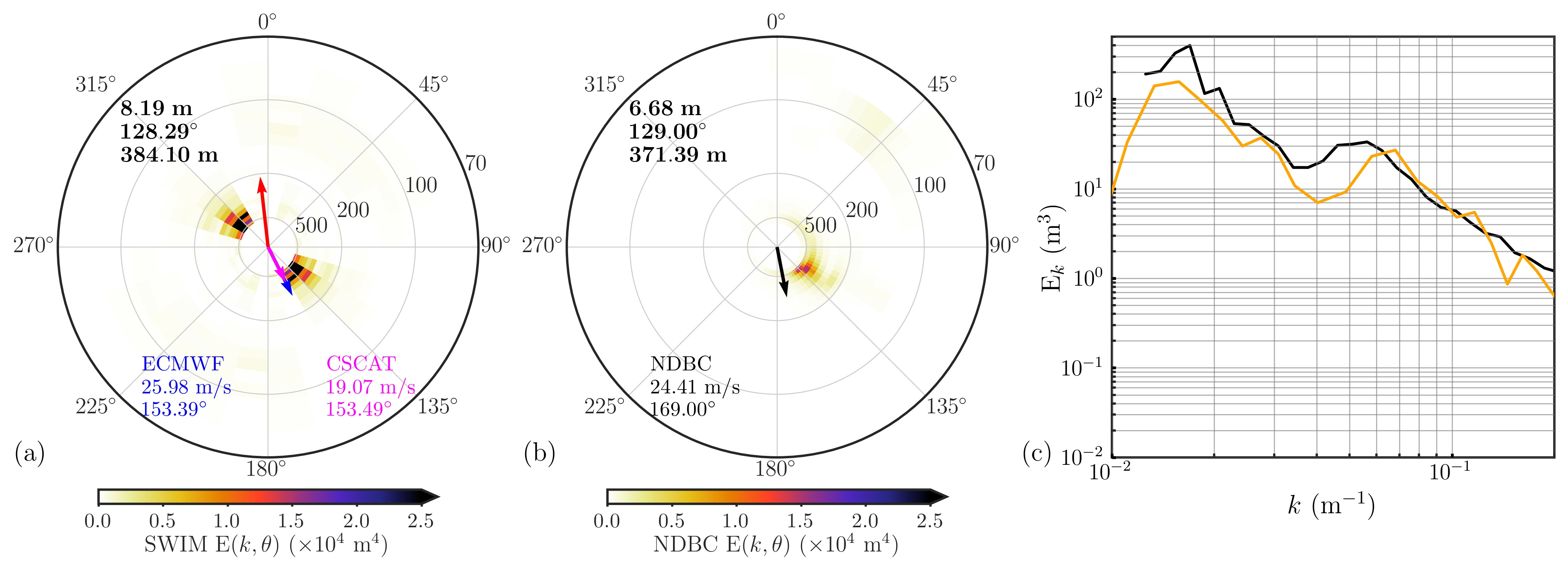}
 \caption{Two-dimensional wave height spectra $\E(k,\theta)$ measured (a) by SWIM in box 78-0 ($10^\circ$ beam) on 1st October 11:32 and  (b)  by the NDBC buoy 41049  on 1st October 11:40. Arrows: wind vectors. (c): Corresponding SWIM (black) and NDBC (orange color) omni-directional wave spectra $\E_k(k)$.}
\label{2D_spectra_SWIM}
 \end{figure}

 To assess for the reliability of SWIM wave spectra under rain conditions, the wave height spectrum of box 78-0 is compared to that of the buoy $41049$, with the distance between the center of the box and the buoy being $33$~km and the time-lapse being only $8$ minutes.  These are the closest SWIM and {\it in situ} measurements available within TC SAM. The wave height spectra are shown in  Figure  \ref{2D_spectra_SWIM}.  Both two-dimensional spectra shown in Figures \ref{2D_spectra_SWIM}a,b exhibit a dominant wave system characterized by a similar dominant wavelength ($384$~m for SWIM and $370$~m for the buoy) coming from the south-east, with $\theta_p\approx 129^\circ$ (with a $180^\circ$ ambiguity for SWIM).  However, the SWIM SWH ($8.2$~m) is overestimated by about $20\%$ compared to the buoy SWH. Either the rain or the fact that the box is closer to the centre of the cyclone by about $30$~km compared to the buoy
could explain this discrepancy. The omni-directional spectra $\E_k(k)$ (in m$^3$), constructed as 
\begin{equation}
\E_{k}= \int_{\theta} \E(k,\theta) \, k \, \dd\theta \,,
\label{Ek}
\end{equation}
are shown in Figure \ref{2D_spectra_SWIM}c. Both profiles are comparable although the energy measured by SWIM at wavelengths larger than $100$~m is larger than that measured by the buoy. A secondary wave system coming from the
northeast is detected by both instruments, at a wavelength of $125$~m according to SWIM and $100$~m according to the NDBC buoy.

\begin{figure}
 \includegraphics[width=1\textwidth]{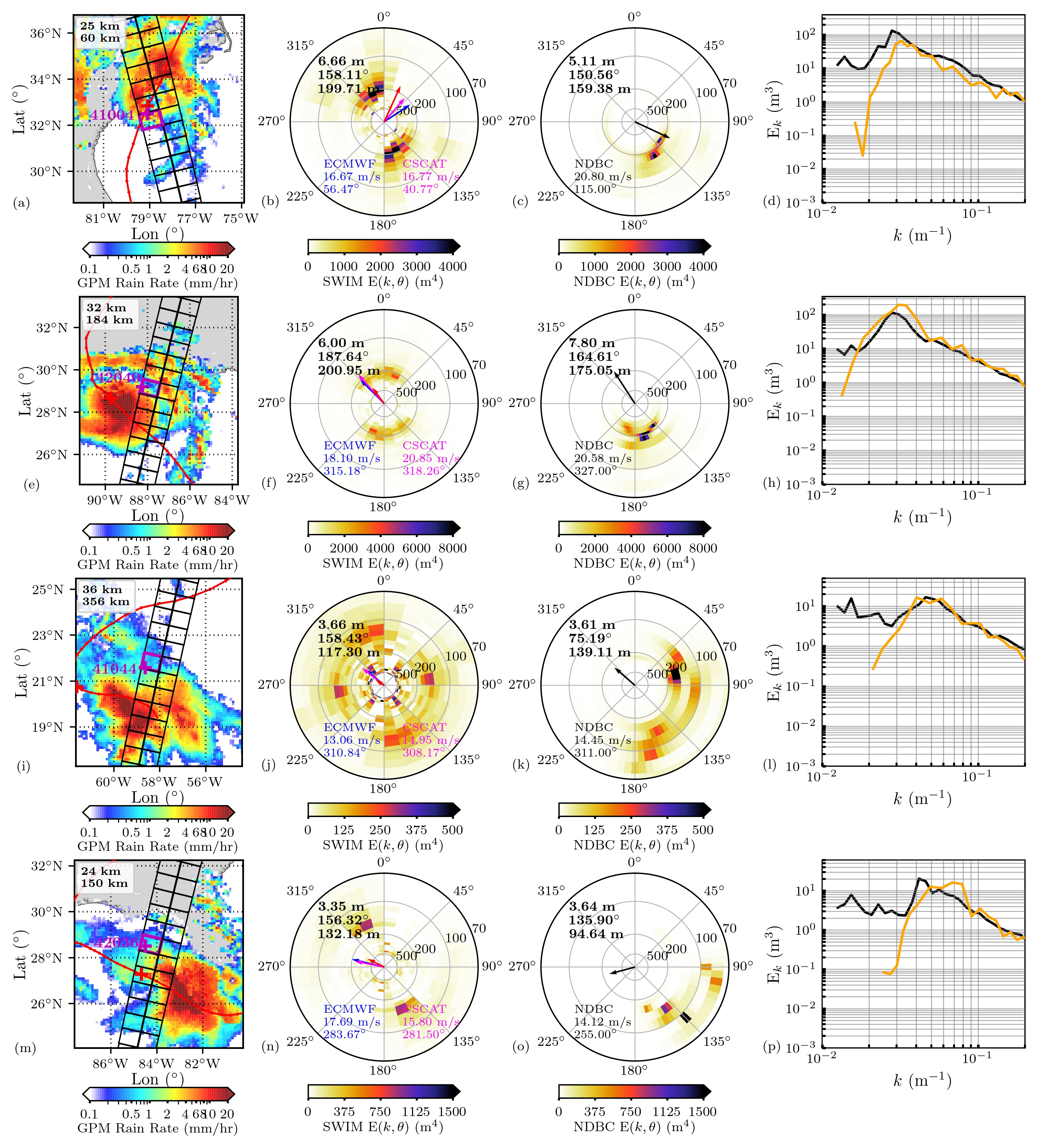}
 \caption{ First column: SWIM boxes (black quadrangles) superimposed to the rain rate field in mm/hr (in colors); red line: the best track of the storm; red cross: storm center during the passage of SWIM; magenta cross: NDBC buoy; white box: SWIM-buoy distance and SWIM-TC distance.
 Second column: SWIM wave height spectrum (closest box to the NDBC buoy, highlighted in magenta in the first column). Third column: NDBC wave height spectrum. Fourth column: 1D wave height spectra. Same as Figure \ref{2D_spectra_SWIM} but for hurricanes Isaias (first row), Ida (second row), tropical storm Sebastien (third row) and hurricane Sally (fourth row).}
\label{OtherTCs1}
 \end{figure}

\begin{figure}
 \includegraphics[width=1\textwidth]{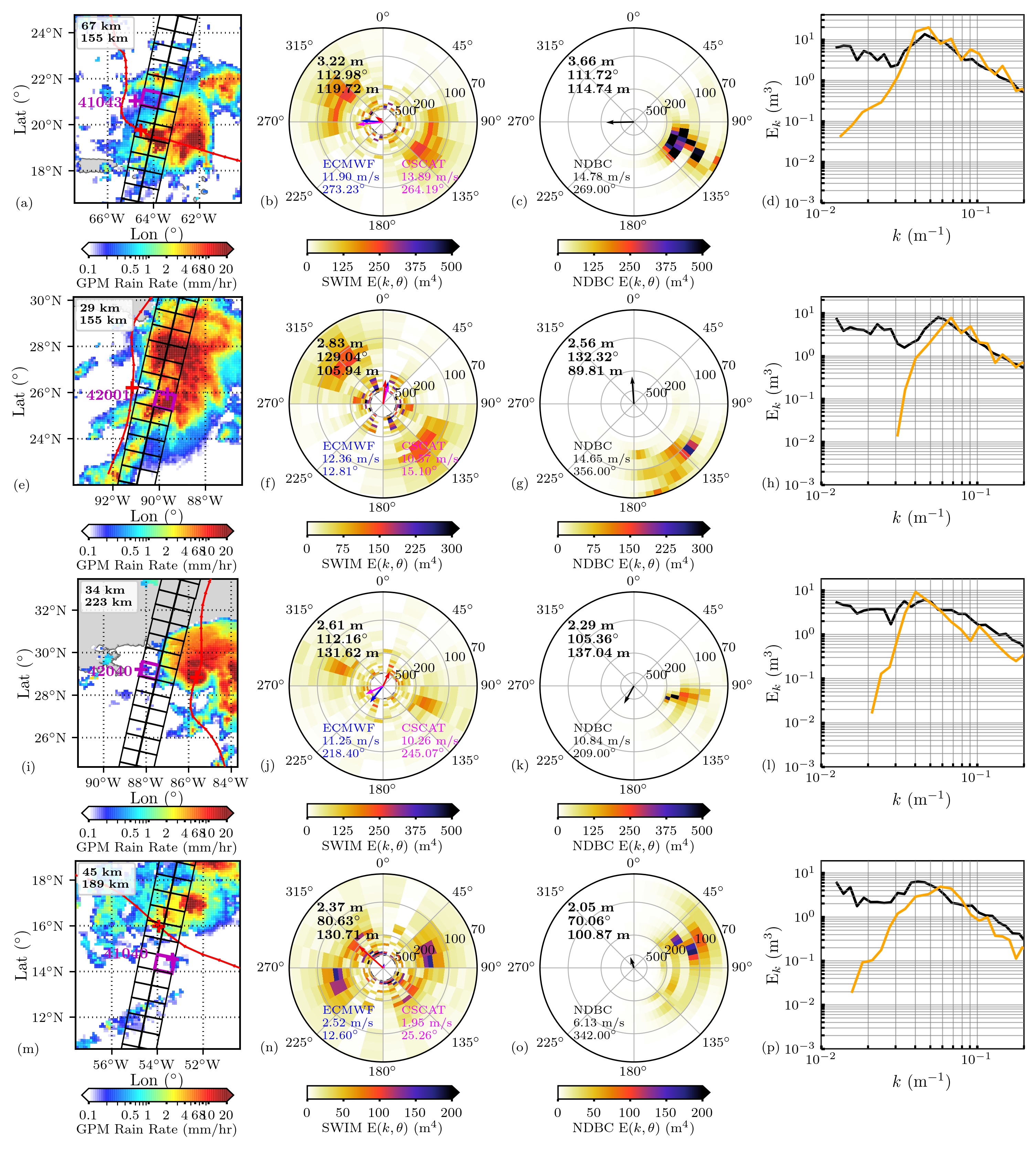}
 \caption{Same as Figure \ref{OtherTCs1}, for hurricane Earl (first row) and tropical storms Claudette (second row), Fred (third row) and Josephine (fourth row).}
\label{OtherTCs2}
 \end{figure}

In the case of SAM, there are unfortunately no other such close comparisons
of satellite and {\it in situ} measurements under high rainfall rates. In
addition to SAM, we however also identified $8$ tropical storms and
hurricanes from 2019 to 2022 for which collocated SWIM and NDBC wave
spectra close to the storm center are available, allowing a maximal distance
between the box center and the buoy of $70$~km. The 2D and 1D wave height
spectra are shown in Figures \ref{OtherTCs1} and \ref{OtherTCs2} (they are
ranked from largest to lowest SWH as measured by SWIM). The maximum rain
rate in the SWIM boxes varies between $0.2$~mm/hr and $48.5$~mm/hr. Even
in the strongest rain cases (first and second row in Figure
\ref{OtherTCs1}, second row in Figure \ref{OtherTCs2}), there is a good
consistency between SWIM and {\it{in situ}} wave spectra, with the wave
parameters differing by less than $20\%$.
The SWIM and {\it in situ} 1D spectra are similar in the peak energy region and for larger values of
  wavenumber. However, they differ significantly at wave numbers lower than $k_p$.
  Indeed, SWIM is known to overestimate energy at small wave numbers, especially for SWH less than about $3$ meters
  \citep[e.g.][]{Jiangetal22}. This effect is clearly visible in the last two spectra shown in Figure~\ref{OtherTCs1}
  and in the four spectra in Figure~\ref{OtherTCs2}.
  In the first case of Figure~\ref{OtherTCs1} (SWH larger than $6$~m), we expect this effect to be less
  pronounced. A similar analysis to that conducted in section~\ref{RainSWIM} was carried out: the energy measured by SWIM at
  low wavenumbers does not appear to be related to rainfall. This case actually corresponds to measurements very close to the
  center of the cyclone (see Figure~\ref{OtherTCs1}a) where spatial variability is significant, which could explain the
  discrepancy between the SWIM and {\it{in situ}} spectra at low wavenumbers. Another possible explanation could be related to
  the choice of the cut-off frequency in the
  post-processing of data acquired by the NDBC buoy, which may  attenuate the energy measured at low
  wavenumbers.

  The results shown in this section exhibit overall consistency between the SWIM and {\it in situ}
  measurements of wave parameters and wave spectra in the wave energy containing part
    (although some artefacts seem to appear at wavenumbers
    much lower than the peak of the spectra), even in the presence of heavy rain.
  Note that the SAR wave spectra acquired in cyclones are more challenging to use. 
  Indeed, these spectra are impacted by the azimuth cut-off effect, which prevents measurements of the wind sea
  waves or short swell. The cut-off is particularly important for wave energy propagating in
  directions close to the SAR azimuth direction. This is highlighted in \ref{appendix_SAR}
  where co-located SWIM and SAR spectra measured within Hurricane SAM are compared
   (see Figure~\ref{SAR1} in \ref{appendix_SAR}).
  Finally, we identified some cases in moderate to heavy rain conditions, where the SAR spectra exhibit a strong
  attenuation compared to SWIM, which can not be attributed to the cut-off effect (because corresponding
  to waves propagating in a direction closer to the range than to the azimuth direction) and which suggest that
  the spectra resulting from the 2D SAR imagettes are more strongly affected by rain than the 1D SWIM footprints (see Figure~\ref{SAR2} in \ref{appendix_SAR}).

\section{Rain impact on SWIM wave spectra}
\label{RainSWIM}

The aim of this section is to understand why the SWIM wave spectra seem reliable, even in heavy rain conditions. 
To do so, we focus on the passage on 1st October 2021 11:32 during hurricane SAM. The footprints corresponding to the spectral beam at $10^\circ$ within $350$~km from the TC center are shown in Figure \ref{Swaths_01oct}a.

\subsection{Rain impact on SWIM radar signals}

\begin{figure}
  \centering
 \includegraphics[width=0.7\textwidth]{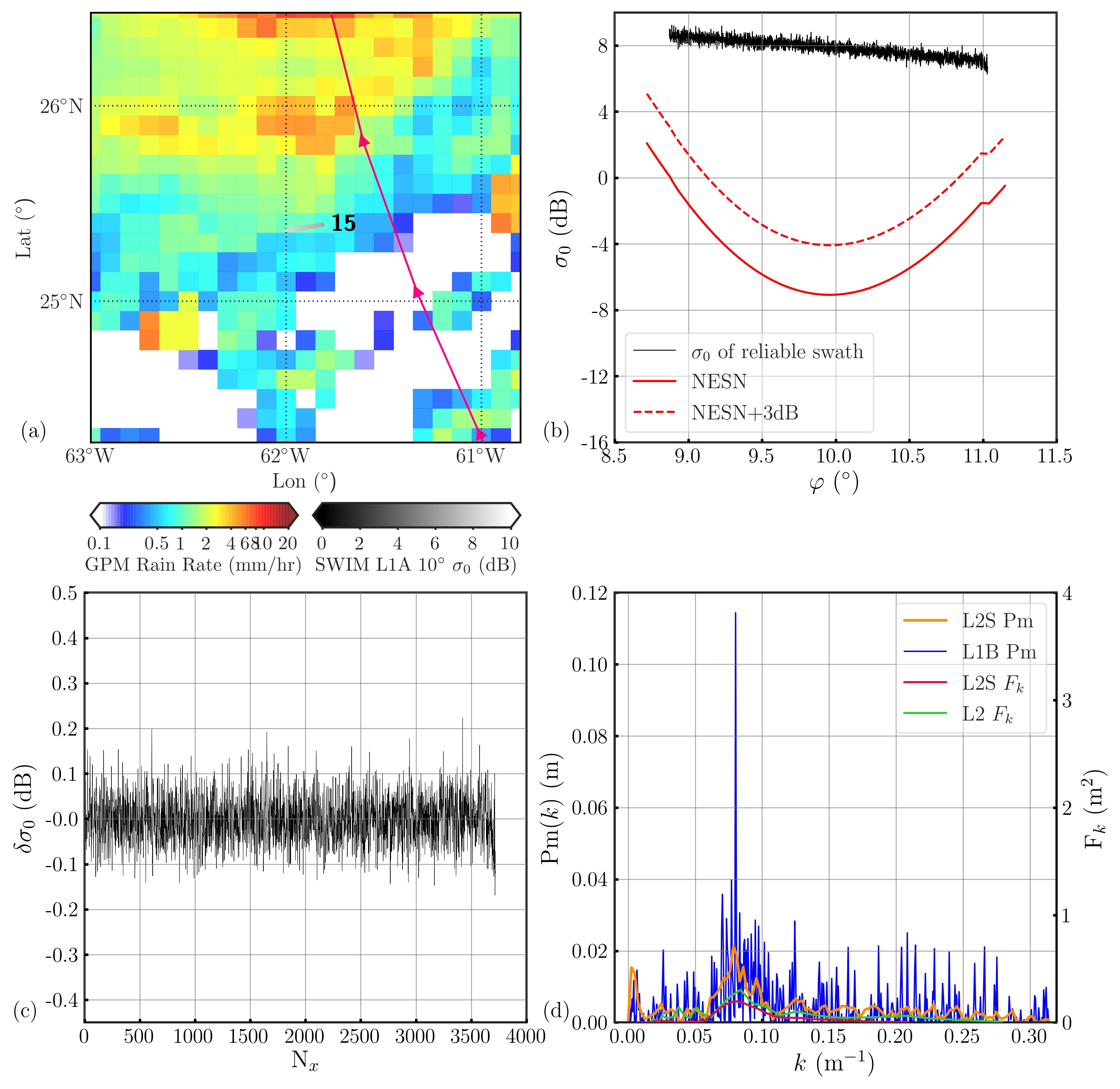}
 \caption{Case of light rain ($1.1$~mm/hr): box 80-1, swath 15, $lsvar=0.0284$, $\phi=80^\circ$.
 (a) Rain rate field (colors) superimposed to the footprint. Dashed: $\sigma_0$ values along the footprint. (b) Black: reliable $\sigma_0$ profile (in dB) as a function of the incidence angle. Solid red: noise equivalent $\sigma_0$. Dashed red: noise equivalent $\sigma_0$ + 3 dB. (c) Fluctuation signal $\delta \sigma_0$ profile (in dB). (d) L1B (blue) and L2S (orange) modulation spectra superimposed to L2 (green) and L2S wave slope spectra.}
 \label{Case1_L1AB}
\end{figure}

\begin{figure}
  \centering
 \includegraphics[width=0.7\textwidth]{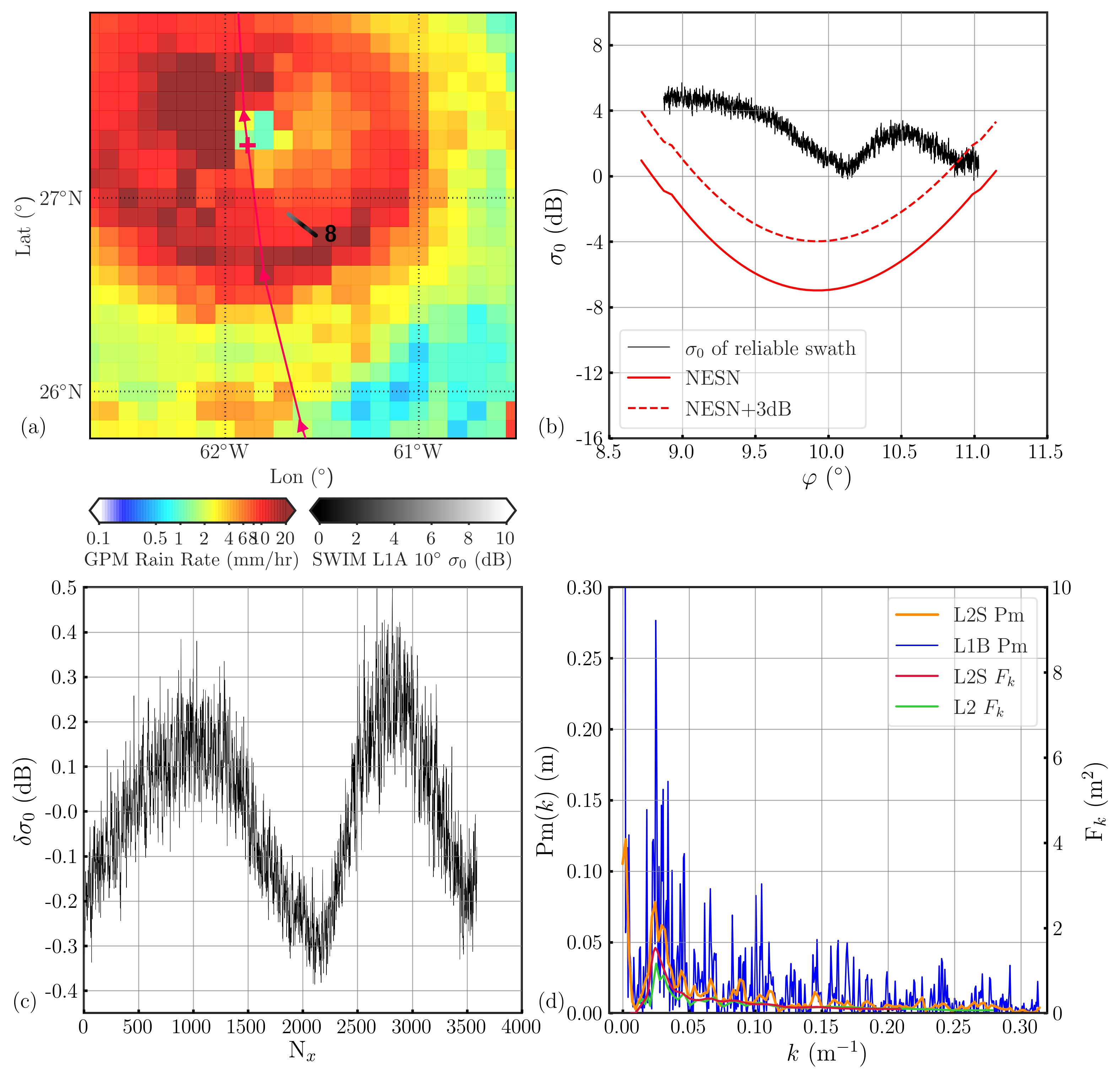}
 \caption{Same as Figure \ref{Case1_L1AB} but in a case of moderate rain ($9.4$~mm/hr): box 78-1, swath 8, $lsvar=0.784$, $\phi=131^\circ$.}
 \label{Case2_L1AB}
\end{figure}

\begin{figure}
  \centering
 \includegraphics[width=0.7\textwidth]{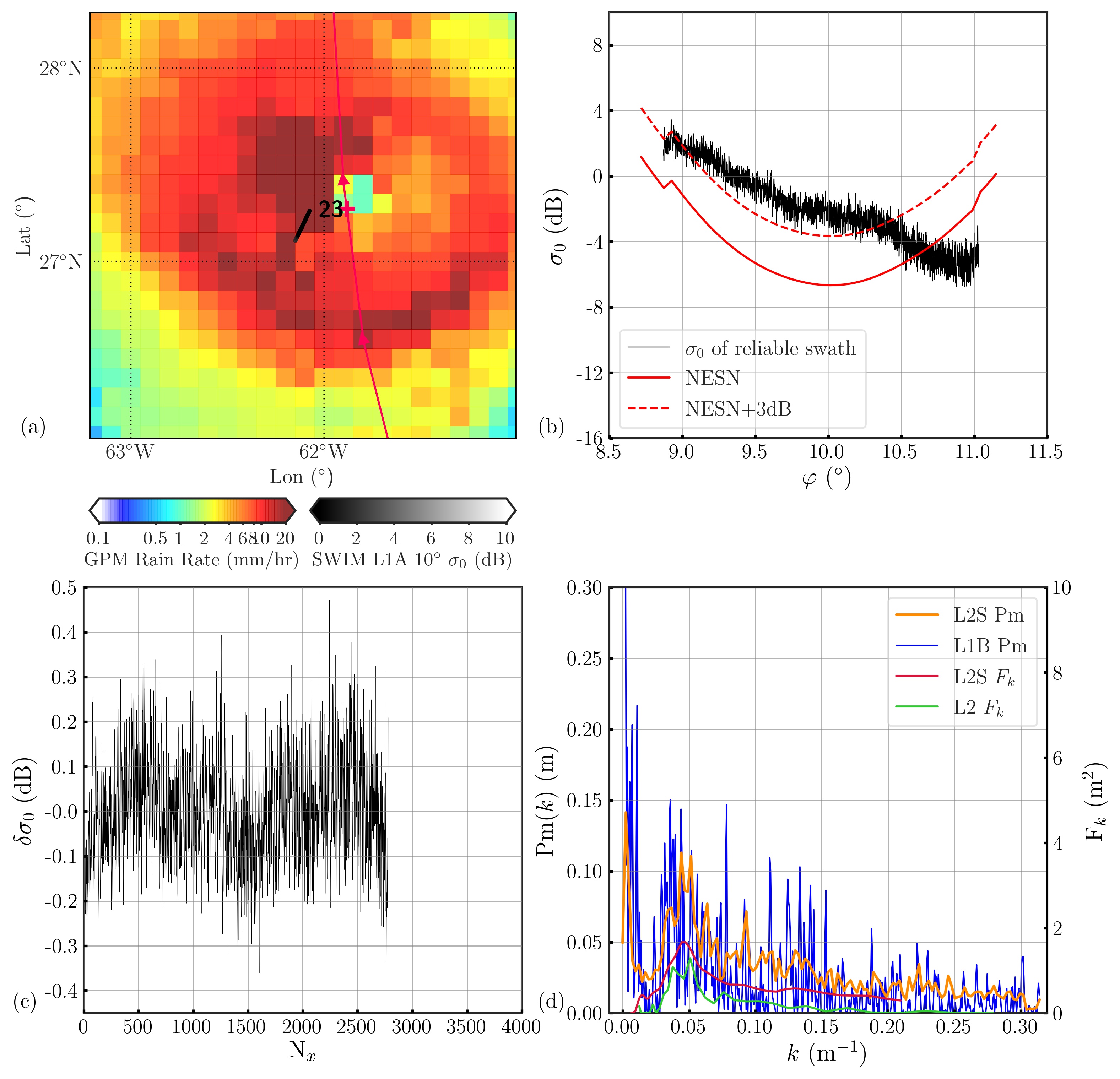}
 \caption{Same as Figure \ref{Case1_L1AB} but in a case of heavy rain ($24.6$~mm/hr): box 78-1, swath 23, $lsvar=0.364$, $\phi=23^\circ$.}
 \label{Case3_L1AB}
\end{figure}

We first focus on three footprints, one in light rain ($1.1$~mm/hr), the other one in moderate rain ($9.4$~mm/hr) and the third one in heavy rain ($24.6$~mm/hr). In following, they are denoted as cases 1, 2 and 3 respectively.  The location of the footprint in the rain rate field is highlighted in Figures~\ref{Case1_L1AB}a, \ref{Case2_L1AB}a and \ref{Case3_L1AB}a, whereas the $\sigma_0$ profile with incidence along the footprint is shown in Figures~\ref{Case1_L1AB}b, \ref{Case2_L1AB}b and \ref{Case3_L1AB}b. The signal is attenuated by rain in cases $2$ and $3$, with the attenuation reaching $12$~dB at the incidence of $11^\circ$ in case $3$ compared to case $1$. Note that the parts of the signal which are too attenuated are filtered in the post-processing. The threshold profile corresponds to the dashed red curve in Figures~\ref{Case1_L1AB}b, \ref{Case2_L1AB}b and \ref{Case3_L1AB}b. Whereas the signal-to-noise ratio is high in case 1, at the far range of the footprint, the signal is close to noise in case 2 and below noise in case 3.
  As far as a significant part of the signal is larger than the threshold, the post-treatment can be conducted.
 For example in case $3$, the signal used as input to the wave
  analysis will be limited to the $9-10.5^\circ$ range in incidence, whereas in case $1$, all values in the
  range $9-11^\circ$ will be kept in the analysis. In case 2, the far range will be limited to about $10.8^\circ$.
 
Additionally, rain inhomogeneities within the footprints can cause distortion of $\sigma_0$. This is particularly visible in case 2 (Figure \ref{Case2_L1AB}b) where $\sigma_0$ exhibits large-scale oscillations at a scale of the order of the swath length (about 20 km for beam 10°), much larger than $500$~m.
The standard deviation of the $\sigma_0$ profile around its mean value, after filtering by a low pass filter (filtering of length scales under $500$~m, associated to waves) allows to quantify the large-scale variability within the footprint. Its value, denoted as $lsvar$, is indicated in the captions of Figures \ref{Case1_L1AB}, \ref{Case2_L1AB} and \ref{Case3_L1AB}: rain inhomogeneities within the footprints, causing large-scale oscillations of the $\sigma_0$ profile, result in larger values of $lsvar$. The signal fluctuations $\delta\sigma_0$ in the L1B product, derived from $\sigma_0$ by subtracting the mean trend, are shown in Figures~\ref{Case1_L1AB}c, \ref{Case2_L1AB}c and \ref{Case3_L1AB}c. As expected, they also exhibit large-scale oscillations due to rain inhomogeneities in cases 2 and 3. 

The resulting modulation spectra, $\Pm(k)$, contained in the L1B (resp. L2S) products are shown in Figures~\ref{Case1_L1AB}d, \ref{Case2_L1AB}d and \ref{Case3_L1AB}d in blue (resp. orange) color lines: the large-scale oscillations of $\delta \sigma_0$ in cases 2 and 3 yield a peak of energy in the modulation spectra at small $k$ in Figures \ref{Case2_L1AB}d and \ref{Case3_L1AB}d. 
It is worth noting that the peak at small $k$ is smaller in the L2S
modulation spectra than in the L1B modulation spectra. Indeed, the L2S
modulation spectra result
from the Welch method \citep{Welch67} where the $\sigma_0$ values are splitted in $15$ overlapping
windows before being used as input of $15$ fast Fourier transforms (FFT) whose spectral energy is then
averaged to build the modulation spectrum. Such a
process logically reduces the impact of the inhomogeneity of rain along the
footprint. In contrast, in the L1B processing, the full length of the $\sigma_0$ samples
  is used as input to the FFT, which leads to more noisy modulation spectra, while the averaging is carried out 
  in the spectral space during the construction of the two-dimensional wave slope spectrum,
  after filtering the low wavenumber components ($k<k_{min}=0.01$~m$^{-1}$). 
Hence, the comparison of L1B and L2S modulation spectra
corroborates the correlation between the large-scale oscillations of
$\sigma_0$ and $\delta \sigma_0$ (as well as the resulting peak
in the modulation spectrum at small $k$) and rain inhomogeneities along the footprint.

The above conclusions have been statistically verified using all the footprints within $350$~km from the TC center during the passage of SWIM on 1st October 2021 (see \ref{appendix_stats}). In \ref{appendix_stats}, using SWIM $10^\circ$ and $8^\circ$ beams, we also show that rain mainly impacts the L1B modulation spectra at small wavenumbers (wavelenghs larger than $1$~km), rather than the waves range ($[70-500]$~m).
A similar study was conducted for the rainiest cases shown in Figures~\ref{OtherTCs1} and \ref{OtherTCs2}, leading to the same conclusions.

\subsection{Rain impact on L2 SWIM wave slope spectra}

The L2 (resp. L2S) wave slope spectra corresponding to the L1B (resp. L2S) modulation spectra are superimposed in Figures \ref{Case1_L1AB}d, \ref{Case2_L1AB}d and \ref{Case3_L1AB}d, in green (resp. red) color lines. They also exhibit a peak of energy corresponding to the waves propagating in the azimuth direction. However, in cases $2$ and $3$, they do not exhibit any peak at small wavenumbers, contrary to the modulation spectra. In the post-processing, values corresponding to wavenumbers smaller than $k_{min}=0.01$~m$^{-1}$ are indeed filtered. Hence, the peak in the modulation spectra at small $k$ due to the large-scale oscillations of $\sigma_0$ does not affect the wave slope spectra. 

\begin{figure}
\centering
 \includegraphics[width=0.7\textwidth]{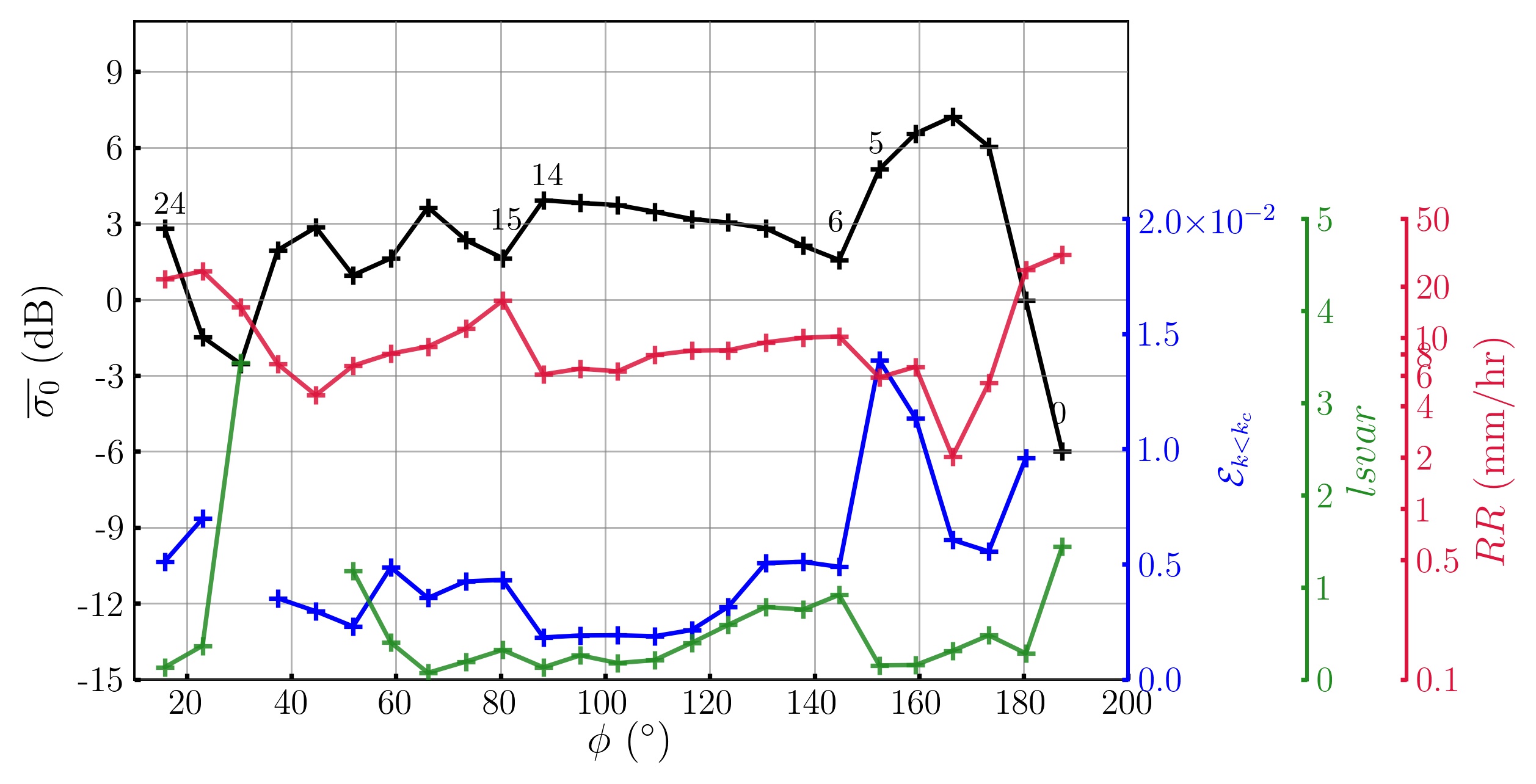}
 \caption{Values of $\overline{\sigma_0}$ (black), $RR$ (red), $lsvar$ parameter (green) and ${\mathcal E}_{k>k_c}$ (blue) as a function of the azimuth direction, corresponding to the $25$ footprints of the $10^\circ$ beam in box 78-1 (see Figure~\ref{fig_swaths_8and10}).}
 \label{fig_box78-1_allswaths}
\end{figure}

As explained in section \ref{DataMethod}, the L2 two-dimensional wave slope spectra at the scale of a box are reconstructed by combining the measurements within the footprints contained in the box.
Figure~\ref{fig_box78-1_allswaths} gathers the averaged parameters of the $25$ footprints of the $10^\circ$ beam in box 78-1 (see Figure~\ref{fig_swaths_8and10}a). The averaged $\overline{\sigma_0}$ is most attenuated around the $30^\circ$ and $190^\circ$ directions, which are characterized by the highest rain rates; this results in large values of the $lsvar$ parameter in these directions. The energy of waves in the modulation spectrum is quantified by ${\mathcal E}_{k>k_c}$ (defined in (\ref{def_E}), with $k_c= 2\pi \cdot 10^{-3}$~m$^{-1}$): it indicates that the dominant direction of waves lies around $150^\circ$. The dominant waves are thus detected by footprints undergoing moderate rain. Generally speaking, it is worth noting that because of the different locations of the footprints in the rain field, different situations are possible. In the configuration where the dominant waves direction coincides with footprints undergoing light or no rain (i.e. the footprints in stronger rain correspond to directions of low energy wave), the rain is not expected to affect the wave spectra, since the wave energy is measured by footprints in light or no rain. In other configurations, where the footprints measuring wave energy undergo moderate or heavy rain (like in Figure~\ref{fig_box78-1_allswaths}), we have shown in this section that as long as the attenuation is not too important, the wave spectra remain reliable.

It is important to note that the post-processing leading to the L2 wave slope spectra involves a
normalization by the nadir SWH. A detailed study of the impact of rain in the post-processing of nadir measurements is out of the scope of the present paper.
However, it can be noted that the comparisons presented in Figures \ref{OtherTCs1}  and \ref{OtherTCs2} do not reveal a clear trend regarding the effect of rain on the nadir SWH.

\section{Wave asymmetry in Hurricane SAM}
\label{Physics}

 Largest waves are found in the right front quadrant of TCs, as highlighted
 in Figure \ref{frame_SAM}a for SAM. The asymmetry in the wave field may
 simply originate from the asymmetry in the wind field (highlighted in
 Figures \ref{frame_SAM}d and \ref{Track_01oct}b), or may be due to
 the so-called extended fetch effect \citep{Young88}. The aim of this
 section is to take advantage of the many measurements available in
 Hurricane SAM to investigate the physical origin of the wave asymmetry.

 The `extended fetch' idea has been recently formalised under
 simplifying assumptions by \cite{Kudry15} here referred to as the KGC15
 model, see also \ref{appendix_KGC15}.
Note that this analytical model has been further extended to a
numerical model \citep{Kudry21a,Kudry21b}. We use here the analytical
version of this work, i.e. that of \cite{Kudry15}.

\begin{figure}
\centering
 \includegraphics[width=1\textwidth]{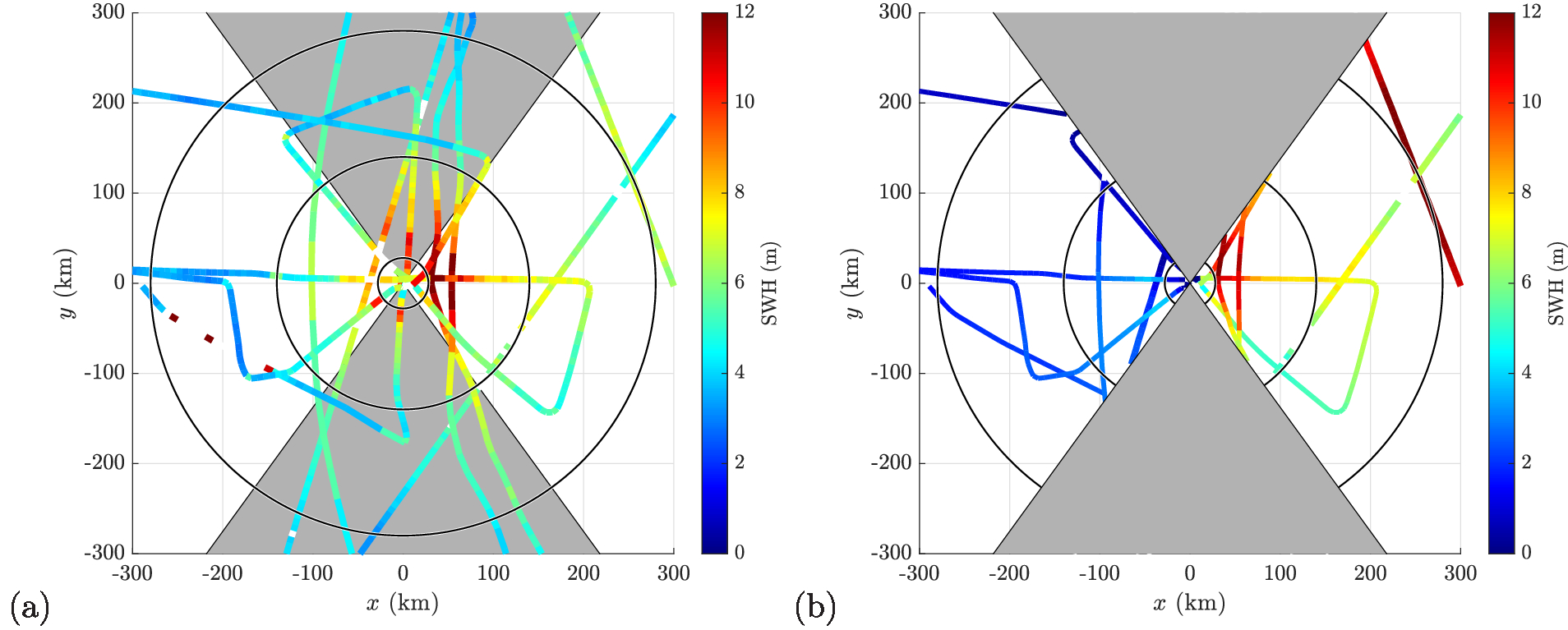}
 \caption{(a) Composite representation of SWH observational data between 29 September 00:00 and 2 October 12:00 reported in the frame of Hurricane SAM. (b) SWH as predicted by the KGC15 model. Shaded gray: area not described by the KGC15 model. Red Black circles indicate $R_{max}$, $5\, R_{max}$ and $10\,R_{max}$.}
 \label{fig_Trapped_SWH}
\end{figure}

The observational data for SWH shown in
  Figure \ref{frame_SAM}a are first compared to the SWH as predicted by the KGC15
model. The latter are calculated by combining (\ref{alpha}), (\ref{Lcr}) and
  (\ref{eqE}) in \ref{appendix_KGC15}.
  For this calculation, we use the surface wind speed measured by the buoys, the SD,
  the CSCAT instrument and the Sentinel 3-SRAL, as well as the ECMWF surface wind speed interpolated
  along the aircraft trajectory. In order to facilitate the comparison,
  the SWH observations are reported in Figure \ref{fig_Trapped_SWH}a, whereas the SWH predicted by the
KGC15 model are juxtaposed in Figure \ref{fig_Trapped_SWH}b.
  The figure is
 presented in the reference frame of the TC and orientated with the TC
 propagation direction upward.
 Note that the wave properties in gray shaded area cannot be
 described by the KGC15 model, because in these areas, the wind vectors are
 nearly orthogonal to the TC propagation. The model reproduces the
 asymmetric wave field. In the right quadrants, the modelled SWH is globally
 consistent with the observed SWH, except on the SWIM track of $2$
 October, for which the model overestimates the SWH.
 This could be explained by the large
   distance of this track from the center of the cyclone (more than $10\, R_{max}$).
 The model also seems to underestimate the SWH in the left quadrants.
 It is important to keep in mind that the KGC15 model is based on strong assumptions,
 including a homogeneous wind field in each of the right and left quadrants. Discrepancies
 between the model estimations
   and the observations are therefore inevitable.
   In particular in the left quadrants the waves in the KGC15 model travel along a direction parallel to the cyclone displacement. No wave energy is radiated from the eyewall in this model. 
   The KGC15 model can however be used to investigate the possible
   existence of trapped waves in the right quadrants of SAM: this is what we tackle in the following.

\begin{figure}
\centering
 \includegraphics[width=1\textwidth]{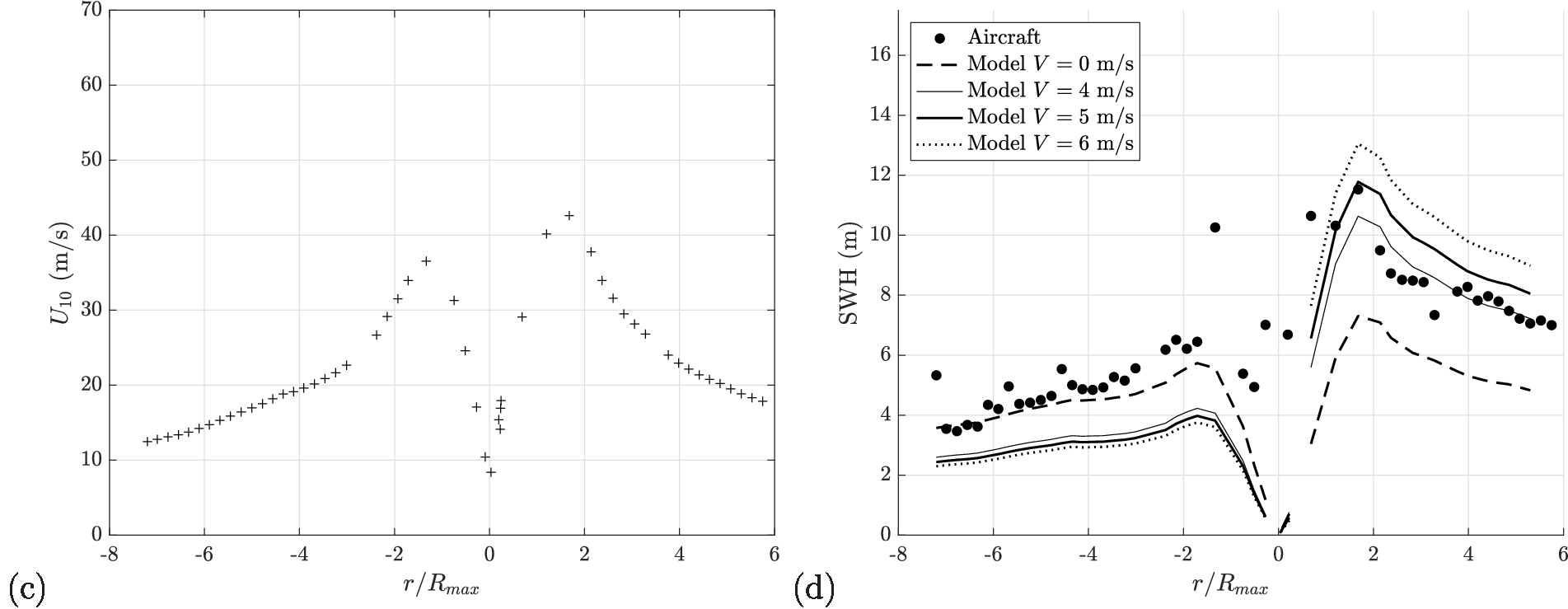}
 \caption{(a) ECMWF wind speed interpolated at the aircraft location, $U_{10}$. (b) SWH as measured by the aircraft (black dots) and estimated using the KGC15 model combined using $U_{10}$ for $V=0$~m$\cdot$s$^{-1}$ (dashed line), $V=4$~m$\cdot$s$^{-1}$ (thin line), $V=5$~m$\cdot$s$^{-1}$ (thick line) and $V=6$~m$\cdot$s$^{-1}$ (dotted line). Abscissa: distance to the TC center (defined with the same sign as $x$) normalized by $R_{max}$.}
 \label{fig_Trapped_AC}
\end{figure}

Since SWIM measurements are either too far from the center of the cyclone or in the region
  not covered by the model, they are no longer used in the following.
Instead, we first focus on the aircraft passage through the right-front
  and left-rear quadrants of SAM
on 30 September between 01:16 and 02:09. The SWH measured by the WSRA onboard the aircraft is compared to 
the modelled SWH, calculated using equation~(\ref{eqE}) in \ref{appendix_KGC15} and the ECMWF
surface wind speed, $U_{10}$, interpolated along the aircraft trajectory (Figure \ref{fig_Trapped_AC}a).
The SWH profiles across SAM are shown in Figure \ref{fig_Trapped_AC}b.
If the TC is assumed to be stationary ($V=0$), the SWH predicted by the KGC15 model strongly underestimates
the SWH observed in the right-front sector (compare the black dots and the dashed line in
Figure \ref{fig_Trapped_AC}b), which suggests that the asymmetry of the wind alone can not explain the observed
asymmetry in SWH.
However, if values of $V$ close to the IBTrACS estimate $V\simeq 5$~m$\cdot$s$^{-1}$ are used, the KGC15 model
allows to retrieve the large SWH observed in the right front sector. That means that a trapped wave mechanism could be
invoked to explain the SWH asymmetry. It is however important to keep in mind that the ECMWF wind
might underestimate winds in SAM. Indeed, during the reconnaissance flight, the
  maximum recorded wind speed in the IBTrACS database is $64$~m$\cdot$s$^{-1}$, which suggests
  that the ECMWF winds shown in Figure \ref{fig_Trapped_AC}a might be slightly underestimated.
We therefore now turn to the
{\it in situ} measurements (saildrone and buoy)
in order to have collocated measurements of waves and wind at the surface.

\begin{figure}
\centering
 \includegraphics[width=1\textwidth]{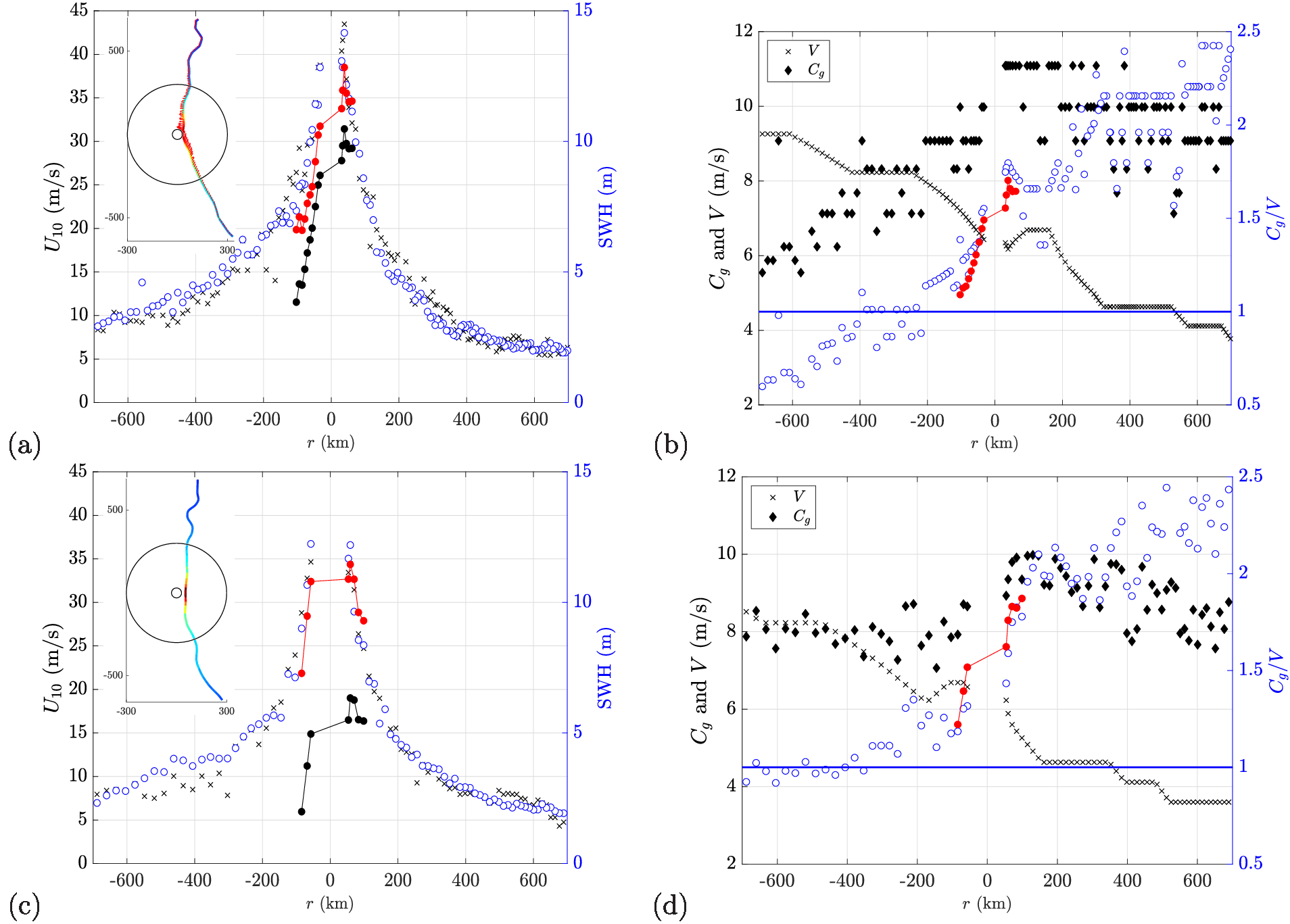}
 \caption{(a) Surface wind speed $U_{10}$ (black crosses) and SWH measured by the SD (blue circles), superimposed to the SWH estimated by the KGC15 model using the translation velocity of SAM $V$ (red line) or setting $V=0$~m$\cdot$s$^{-1}$ (black line). (b) Group velocity $C_g$ measured by the SD (black diamonds) and $V$ (black crosses), superimposed to $C_g/V$ (blue circles) and to its estimation by the KGC15 model (red line). Blue line indicates $C_g/V=1$. (c) and (d): Same as (a) and (b) but for the buoy $41044$. Abscissa: distance to the TC center (defined with the same sign as $y$). The trajectory in the frame of SAM of the SD and of the buoy $41044$ (colored by the SWH and superimposed to the wind vectors) are shown on the insets in panels (a) and (c).}
 \label{fig_Trapped_Insitu}
\end{figure}

This issue is thus finally addressed in Figure \ref{fig_Trapped_Insitu} using
the {\it in situ} measurements by the SD and by the buoy 41044.
Indeed, these two instruments provide measurements of wind and waves across the right
  quadrants of SAM, where the phenomenon of trapped waves is likely to occur,
  passing within a few tens of kilometers from its center
  (see insets in Figures~\ref{fig_Trapped_Insitu}a and \ref{fig_Trapped_Insitu}c).
The surface wind speed and SWH measured by the SD from 29 September 01:00 to 1st
October 17:30 (resp. by the buoy $41044$ from 28 September 10:40 to 1st
October 10:40) are shown in Figure \ref{fig_Trapped_Insitu}a (resp. Figure
\ref{fig_Trapped_Insitu}c), as a function of the distance to the TC center,
$r$, defined as positive in the right-front quadrant and negative in the
right-rear quadrant. The SWH as estimated by the KGC15 model using the {\it
  in situ} wind profiles as well as the propagation velocity $V$ of SAM
(shown in Figure \ref{fig_Trapped_Insitu}b and Figure
\ref{fig_Trapped_Insitu}d)  is superimposed in red in Figure \ref{fig_Trapped_Insitu}a and 
Figure~\ref{fig_Trapped_Insitu}c. Setting $V=0$ in the
model yields the black curve. In both cases, the latter strongly
underestimates the SWH, whereas taking account the motion of SAM yields
values much closer to the SWH measurements. This implies that the large
values of SWH observed in the right quadrants of SAM by the SD and by the
buoy $41044$ are consistent with the extended fetch theory.
Note that the critical fetch defined in (\ref{Lcr}) for winds blowing at $40$~m$\cdot$s$^{-1}$
  and a translation velocity $V=7$~m$\cdot$s$^{-1}$ (which is the value of $V$ according to IBTrACS
  when both devices are closest to the center of SAM) is approximately $30$~km. This corresponds
  to a distance of about $10$~km from the axis.
  The SD and the buoy 41044 thus operate in regions where conditions are favorable for trapped waves.
To put it further to the test, the group velocity $C_g$ is calculated from the peak
period $T_p$ measured by both devices using the gravity waves dispersion
relation. It is shown in Figure \ref{fig_Trapped_Insitu}b and Figure
\ref{fig_Trapped_Insitu}d (see the black diamonds).  It can be noticed that the group velocity
measured by the SD exhibits more variability than that measured by the buoy, due to
fluctuating values of $T_p$.  
In the KGC15 model, which involves a constant $V$, trapped waves are
characterized by $C_g = V$ at the turning point, where waves change
direction (from backward to forward), and by an increasing $C_g$ as they
move forward. This tendency is consistent with the increasing dominant wavelength as the
SD and the buoy 41044 move forward to the front of SAM (Figure \ref{frame_SAM}b).
The propagation speed $V$ of SAM during the time period
investigated here keeps decreasing as $r$ decreases (see the black crosses
in Figures \ref{fig_Trapped_Insitu}b and \ref{fig_Trapped_Insitu}d). In
order to get rid of this tendency, the ratio $C_g/V$ is superimposed to the
$C_g$ and $V$ profiles in Figures \ref{fig_Trapped_Insitu}b and
\ref{fig_Trapped_Insitu}d (blue dots). The ratio $C_g/V$ as measured by the
SD (resp. by the buoy $41044$) varies from $0.5$ (resp. $1$) to $2.5$ as
$r$ increases, which is consistent with growing waves propagating
forward. The region where $C_g$ is approximately equal to $V$
  (compare the blue dots to the blue line indicating $C_g=V$ in
  Figures~\ref{fig_Trapped_Insitu}b and \ref{fig_Trapped_Insitu}d) is located upstream of
  the region with the highest SWH
  (blue dots in Figures~\ref{fig_Trapped_Insitu}a and \ref{fig_Trapped_Insitu}c),
  which is expected in the
trapped wave phenomenon.
The ratio $C_g/V$ as predicted by the KGC15 model is superimposed (see the red curve
in Figures~\ref{fig_Trapped_Insitu}b and \ref{fig_Trapped_Insitu}d):
it provides a good description of the $C_g/V$
profile, which corroborates a wave trapping mechanism in the area explored
by the SD and the buoy $41044$. 

\section{Conclusions}
\label{Conclusion}

The present paper takes advantage of the numerous observational data in
Hurricane SAM (2021) from satellite, aircraft and {\it in situ}
instruments, to investigate the physics of waves in Hurricane SAM. 
Whereas the SWIM wave measurements have been used in several previous
studies to investigate waves in tropical cyclones, the relevance of the
wave spectra in such heavy rain conditions has not been investigated so
far. 
In the present study, we assess the reliability of SWIM measurements in
TCs, by comparing the SWIM wave spectra, as well as the averaged parameters
(SWH, dominant wavelengh, dominant direction), to that measured by $3$ NDBC
buoys and a saildrone (SD 1045) close to the track of Hurricane
SAM. According to the spatial and temporal variability of waves in TCs,
such a comparison makes sense only if the satellite and {\it situ}
measurements are close enough, both in time and space, which strongly
restricts the number of relevant comparisons. The case of Hurricane SAM,
was thus complemented by $8$ additional storms and hurricanes, for which
collocated SWIM and NDBC measurements were performed in the vicinity of the
storm, in various rain conditions, including heavy rain conditions. Those
comparisons confirm the reliability of SWIM wave spectra, even in heavy
rain conditions, despite a tendency for
  overestimating the energy at small $k$ (more pronounced
for SWH less than about $3$ meters) in the presence or absence of rain \citep{Jiangetal22}.

To understand why the SWIM spectra remain reliable in heavy rain conditions,
  we then focused on the post-processing chain of the SWIM
instrument, from the radar backscattering signal to the 2D wave spectra,
which are reconstructed by combining the measurements of wave energy
propagating along footprints covering the angular range $\left[0,
  180\right]^\circ$. We have shown that heavy rain yields an attenuation of
the normalized radar cross section $\sigma_0$ and that in case of rain
inhomogeneity along the footprints, the $\sigma_0$ signal is distorted,
leading to an anomalous peak at small wavenumber (wavelengths larger than
$1$~km) in the modulation wave spectra. However, thanks to a filtering of
both the values corresponding to a low signal-to-noise ratio and the energy
at wavenumbers smaller than $k_{min}=0.01$~m$^{-1}$, both effects do not
affect the resulting 2D wave spectra.

We then investigated the existence of
trapped waves in Hurricane SAM, using a combination of satellite, aircraft and {\it in
  situ} wind and wave measurements. To do so, we also used the analytical
model of \cite{Kudry15} which predicts the wave properties (SWH and peak
wavelengh) in the right and left quadrants of a moving TC, given a surface
wind speed profile.
The model relies on the idea that in the right quadrants of TCs,
some waves may experience an extended fetch due to the displacement of the
TC. This reinforces the SWH asymmetry induced by the wind speed asymmetry
and results in the highest waves on the right of the TC track in the North Hemisphere.
Using the wind and wave measurements by a SD and a NDBC buoy through the
right quadrants of hurricane SAM, we have shown that a trapped wave
mechanism may be invoked to explain the large SWH measured by both devices
in the right-front quadrant.
It is also consistent with the aircraft measurements,
although significant uncertainties remain regarding the ECMWF wind speed.
Note that the orientation and the distance of the SWIM tracks through hurricane SAM were
not optimal for comparisons with the model.  

Beside the impressive images and movies provided by the saildrones in TCs,
their wind and wave measurements can be very valuable to improve our understanding
of the generation of waves in TCs. The same applies to the wave
measurements by SWIM whose wave spectra have been shown, in the present
study, to be reliable even under heavy rain conditions.
With more saildrones being deployed, and as SWIM continues to operate,
further hurricanes will be studied using the combined approach advocated in
this paper and will most certainly improve our understanding of the most
threatening waves under a tropical cyclone.

\section{Open Research}

The CFOSAT SWIM data are available on {https://aviso-data-center.cnes.fr} (L1A) and {https://www.aviso.altimetry.fr/en/data/products/wind/wave-products/wave-wind-cfosat-products.html} (L1B and L2). The SWIM L2S data are provided by the Ifremer Wind and Wave Operation Center on {https://cersat.ifremer.fr/fr/Projects/Recent-and-ongoing-projects/IWWOC}. The {\it in situ} data are available on the NOAA website:
{https://data.pmel.noaa.gov/pmel/erddap/tabledap/sd1045\_hurricane\_2021.html} and {https://www.ndbc.noaa.gov}\,.
The IBTrACS database can be downloaded from { https://www.ncei.noaa.gov/products/international-best-track-archive-for-climate-stewardship-ibtracs/v04r00}. The ECMWF wind data are available on the ECMWF's Meteorological Archival and Retrieval System ({https://confluence.ecmwf.int/display/CEMS/MARS}). Finally, the GPM IMERG precipitation data can be downloaded from NASA {https://gpm.nasa.gov/data/directory}.

\section*{Acknowledgments}

This work was performed during a research visit of Xiaolu Zhao in Sorbonne Université from April 2022 to April 2024. This visit was supported in part by the China Scholarship Council (CSC) Ph.D. Joint Training Program under Grant \#202109040021. This research project is partly supported by the CNES/TOSCA (Maeva project). We thank the NOAA and the Saildrone company for sharing the saildrone data. 

\clearpage

\appendix
\section{SAR performances}
\label{appendix_SAR}

\begin{figure}
\centering
 \includegraphics[width=0.7\textwidth]{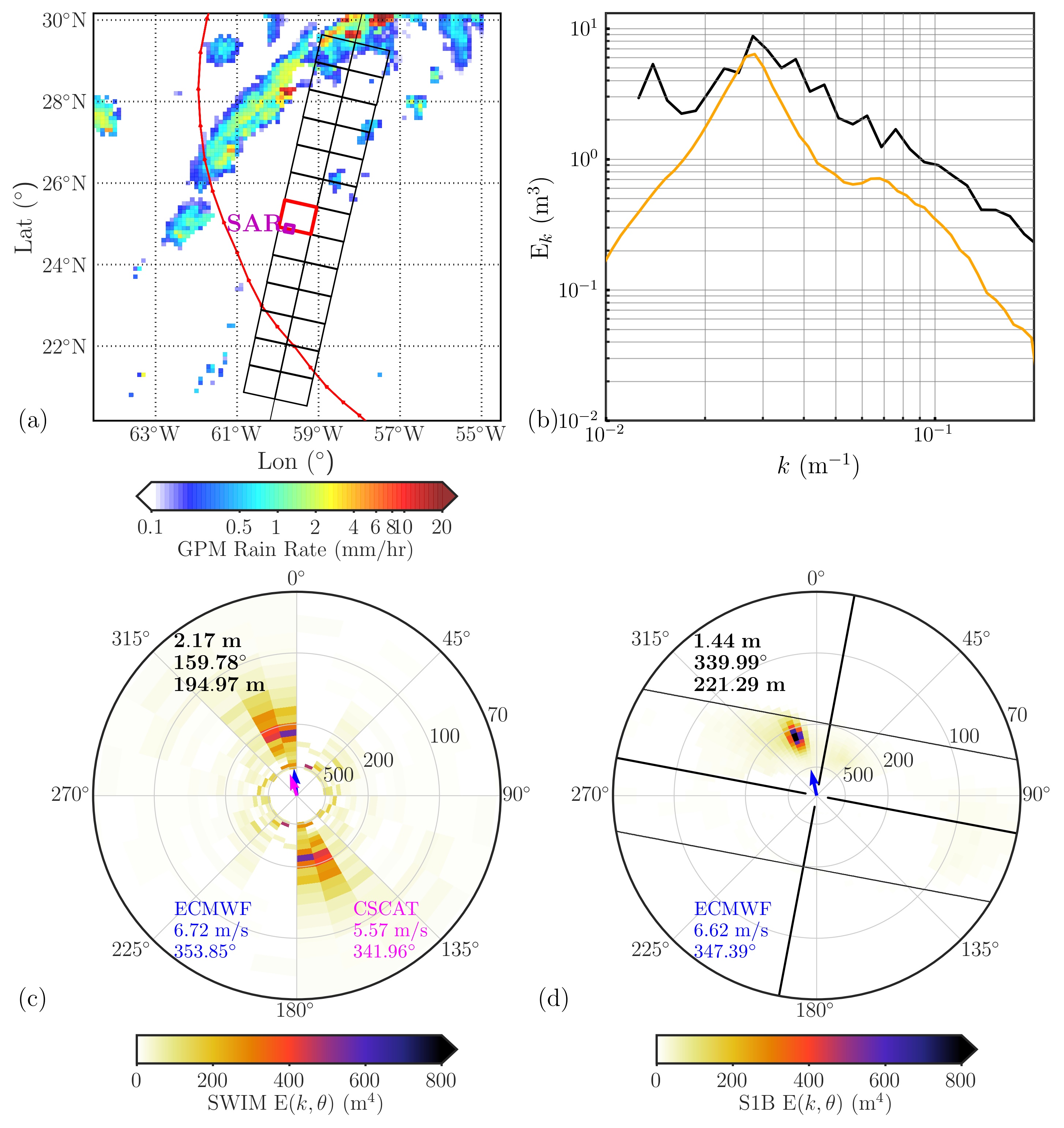}
 \caption{(a) SWIM boxes (black quadrangles) and SAR imagette (in purple) superimposed to the rain rate field in mm/hr (in colors); red line: the best track of the storm; red cross: storm center during the passage of SWIM.
 (b)  SWIM (black) and SAR (orange) 1D wave height spectra corresponding to the SWIM box highlighted in red in (a) and to the SAR imagette. Corresponding (c) SWIM and (d) SAR 2D wave height spectra; thick solid lines: azimuth and range directions, thin solid line: shortest detectable wavelengths.}
\label{SAR1}
\end{figure}

\begin{figure}
\centering
 \includegraphics[width=0.7\textwidth]{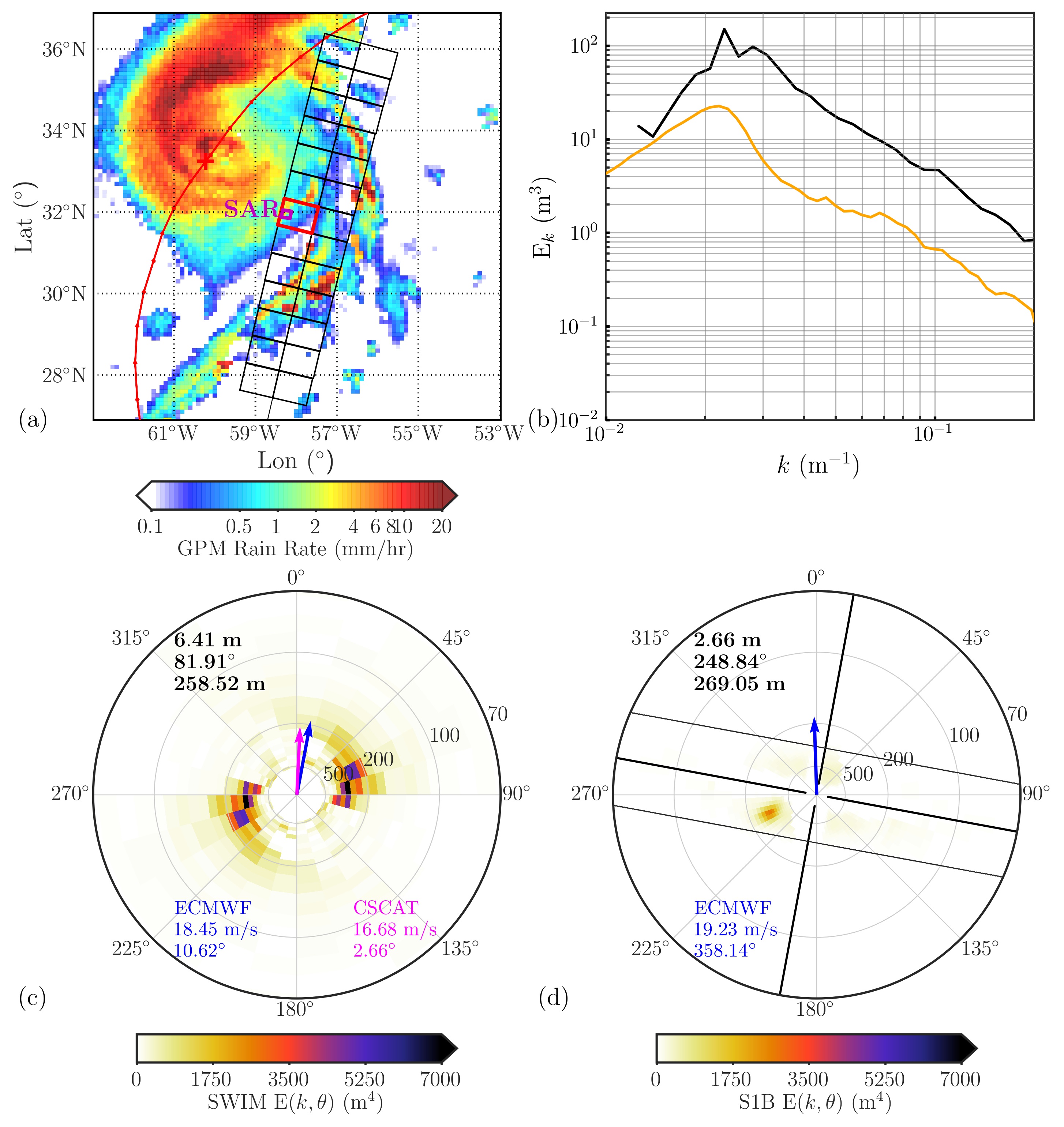}
 \caption{Same as Figure~\ref{SAR1} but corresponding to SWIM boxes and SAR imagette under
     moderate rain conditions.}
\label{SAR2}
\end{figure}

On 2 and 3 October 2021, the SWIM-CFOSAT and SAR-Sentinel1 instruments passed closed to each other
(SWIM boxes and SAR imagettes being located at less than $40$~km)
in the rainbands of SAM with an approximate time-lapse of about 1h30.
  Figure \ref{SAR1} shows a first example of collocated SWIM and SAR wave height spectra on 2 October.
Their locations are highligted by the red and purple rectangles in Figure~\ref{SAR1}a, superimposed
to the rain rate field: in this example no rain is detected within the SWIM box and the SAR imagette.
The SAR SWH is a factor $1.5$ smaller than the SWIM nadir SWH, and the SAR detects much less energy than SWIM at
wavenumbers larger than $0.03$~m$^{-1}$ (i.e. wavelenghs smaller than $160$~m): this attenuation is
characteristic of the cut-off effect. A second comparison, this time under moderate rain,
is carried out in Figure~\ref{SAR2}. In this case,  
the rain rate reaches $3.4$~mm/hr in the SWIM box and $3.2$~mm/hr in the SAR imagette.
The SAR SWH is smaller by a factor $2.5$ compared to the SWIM nadir SWH and the SAR spectrum is strongly
attenuated over the entire wavenumber range, compared to SWIM.
Since the waves propagate in a direction closer to the range direction than to the azimuth direction,
the cut-off effect can not explain such a discrepancy between both spectra and the latter is more
likely due to rain.

\clearpage

\section{Statistical study of the rain impact on modulation wave spectra}
\label{appendix_stats}
Beyond the SWIM measurements along the $3$ footprints shown in Figures \ref{Case1_L1AB}, \ref{Case2_L1AB} and \ref{Case3_L1AB}, the measurements within the footprints within $350$~km from the TC center during the passage of SWIM on 1st October 2021 are now investigated.

\begin{figure}
\centering
 \includegraphics[width=0.4\textwidth]{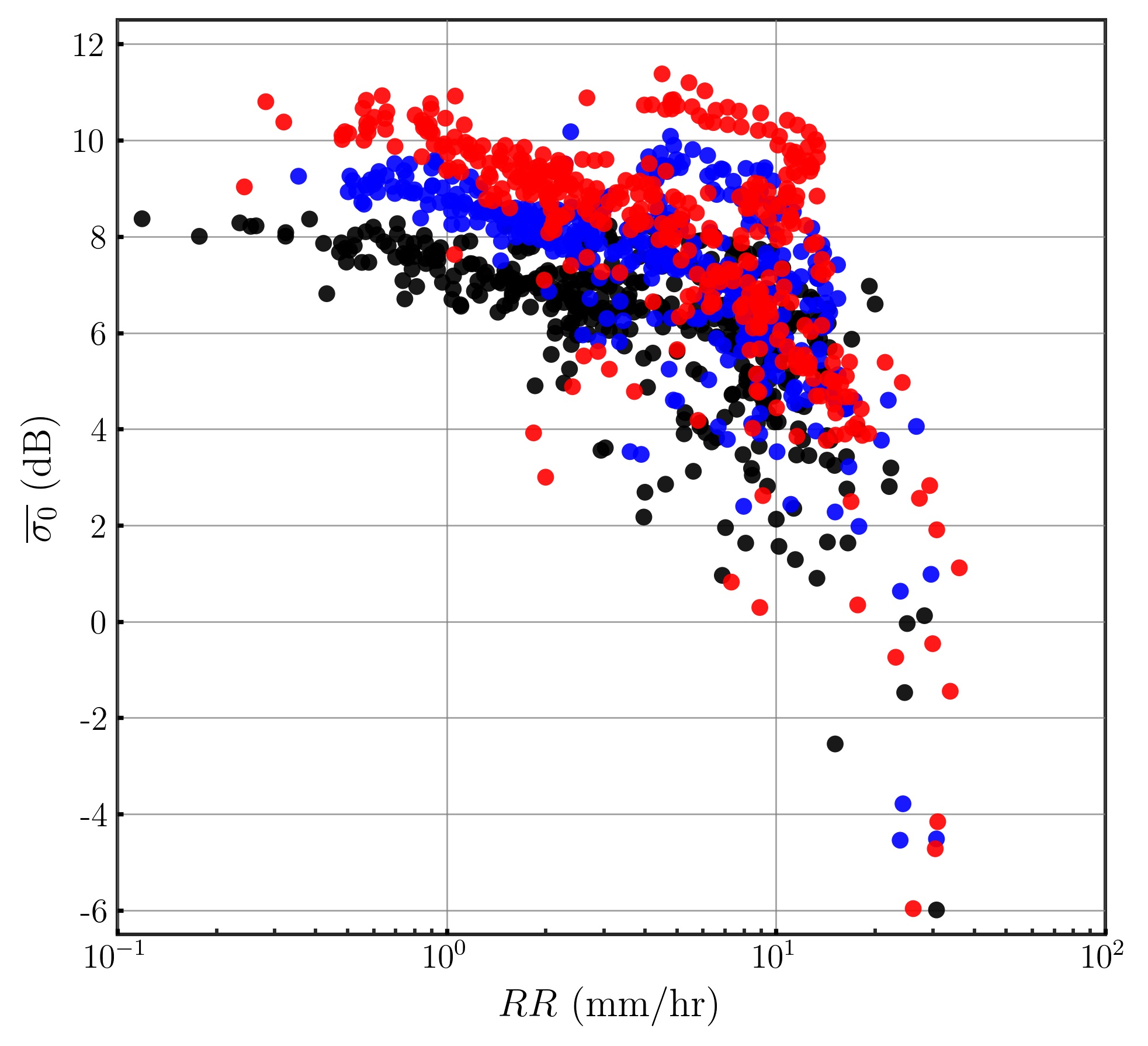}
 \caption{Averaged value of the reliable $\sigma_0$ profile ($\overline{\sigma_0}$) as a function of the collocated rain rate ($RR$) estimated at the middle of the $405$ footprints of the $6^\circ$ beam (red color) and  $407$ footprints  of the $8^\circ$ (blue color) and $10^\circ$ (black color) beams. }
\label{sigma0_rain}
\end{figure}

\begin{figure}
\begin{center}
 \includegraphics[width=0.9\textwidth]{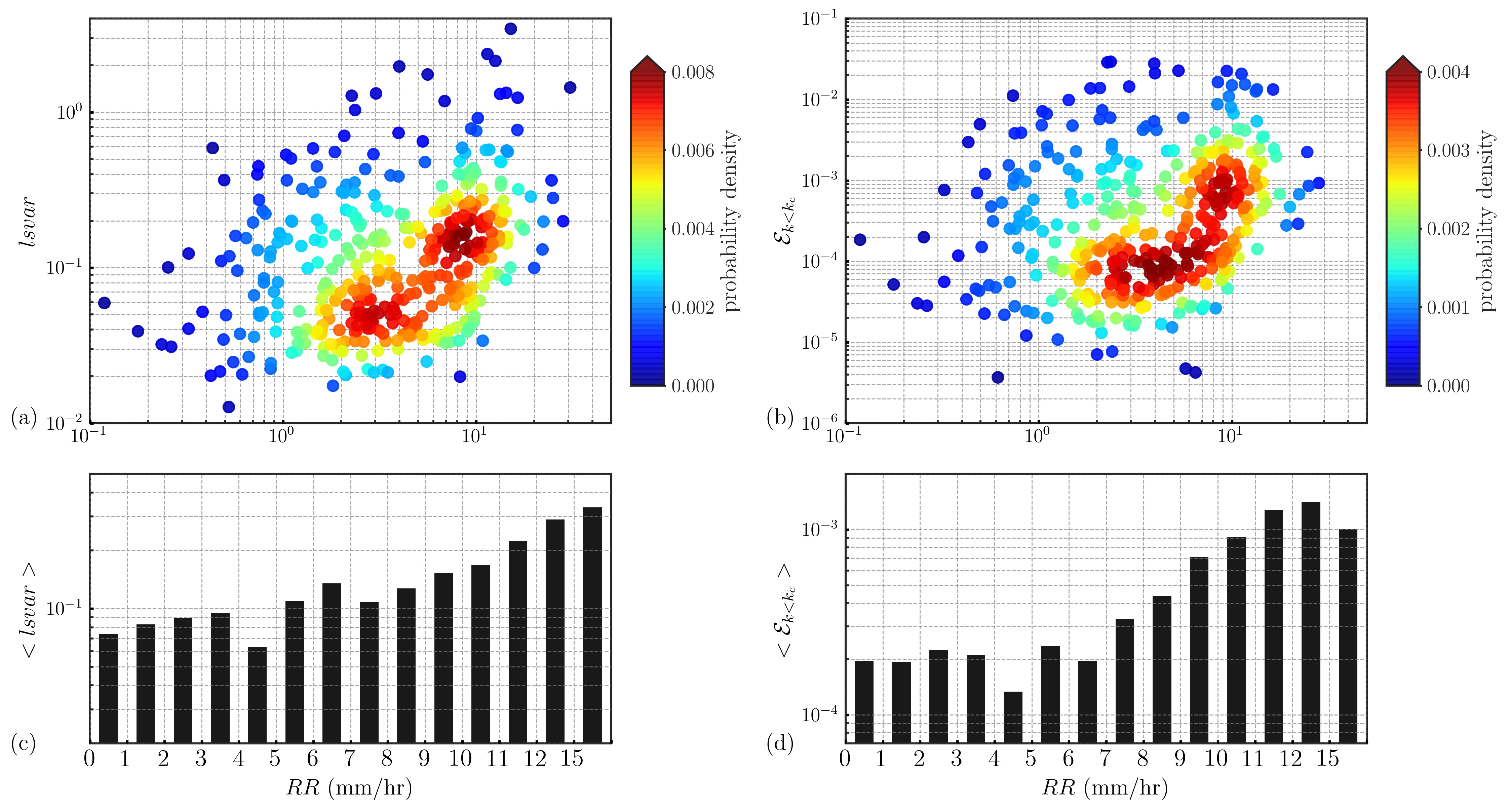}
 \end{center}
\caption{Scatterplot and probability density function (colors) of (a) the large-scale variability within the footprints ($lsvar$) and of (b) the energy ${\mathcal E}_{k < k_c}$, as a function of the rain rate $RR$ at the middle of the $407$ footprints of the $10^\circ$ spectral beam. Corresponding averaged values of (c) $lsvar$ and of (d) ${\mathcal E}_{k < k_c}$ per category of $RR$.
}
 \label{lsvar_Pmsmallk_rain}
\end{figure}

Figure \ref{sigma0_rain} shows the averaged value of the reliable $\sigma_0$ profiles ($\overline{\sigma_0}$) as a function of the collocated rain rate ($RR$) estimated at the middle of the footprints, for the $3$ spectral beams. It highlights the attenuation of $\overline{\sigma_0}$ by rain. For a $RR$ of $10$~mm/hr, it reaches up to 6 dB for the $10^\circ$ beam. 
The spatial resolution of the IMERG product (about $10$~km) is unfortunately too low for quantifying the rain inhomogeneities along the footprints (about $20$~km length). The following figures thus involve the rain rate at the middle of the footprints. 
Figure \ref{lsvar_Pmsmallk_rain}a shows the large-scale variability of 
$\sigma_0$ within the footprints of the $10^\circ$ spectral beam (quantified by the $lsvar$ parameter) as a function of the rain rate $RR$. For moderate and heavy rains, there is a correlation between $lsvar$ and $RR$. The representation of the averaged value of $lsvar$ per rain rate category in Figure~\ref{lsvar_Pmsmallk_rain}c corroborates the correlation between both quantities when $RR$ is larger than $7$~mm/hr. The energy of the small $k$ peak on the modulation spectra is quantified by 
\begin{equation}
{\mathcal E}_{k < k_c}=\int_{k < k_c}\Pm (k)dk \,,
\label{def_E}
\end{equation}
where $k_c=2 \pi \cdot 10^{-3}$~m$^{-1}$ (i.e. $\lambda_c=1$~km). This quantity corresponds to the energy of the modulation spectra at wavelengths larger than $1$~km. This threshold value was chosen in order to distinguish the small $k$ anomalous peak to the ocean wave spectrum. Figure \ref{lsvar_Pmsmallk_rain}b shows ${\mathcal E}_{k < k_c}$ as a function of $RR$, whereas its averaged value per rain rate category is represented in Figure \ref{lsvar_Pmsmallk_rain}d. These figures highlight the correlation of ${\mathcal E}_{k<k_c}$ with $RR$ for moderate and heavy rains.
Actually, the rain rate at the middle of the footprint is not necessarily correlated to the rain inhomogeneities along the footprint. For example, the rain rate in case 3 (Figure \ref{Case3_L1AB}) is larger than that in case 2 (Figure \ref{Case2_L1AB}). Yet, the $\sigma_0$ signal is more distorted in case $2$, suggesting that rain inhomogeneities are more important in case 2. This can explain why the correlations shown in Figure \ref{lsvar_Pmsmallk_rain} are not stronger. The rain product spatial resolution (about $10$~km) is definitely a limiting factor in the investigation of the impact of rain inhomogeneities on SWIM measurements. This is also the case for its half-hourly temporal resolution. 

\begin{figure}
\centering
\includegraphics[width=0.8\textwidth]{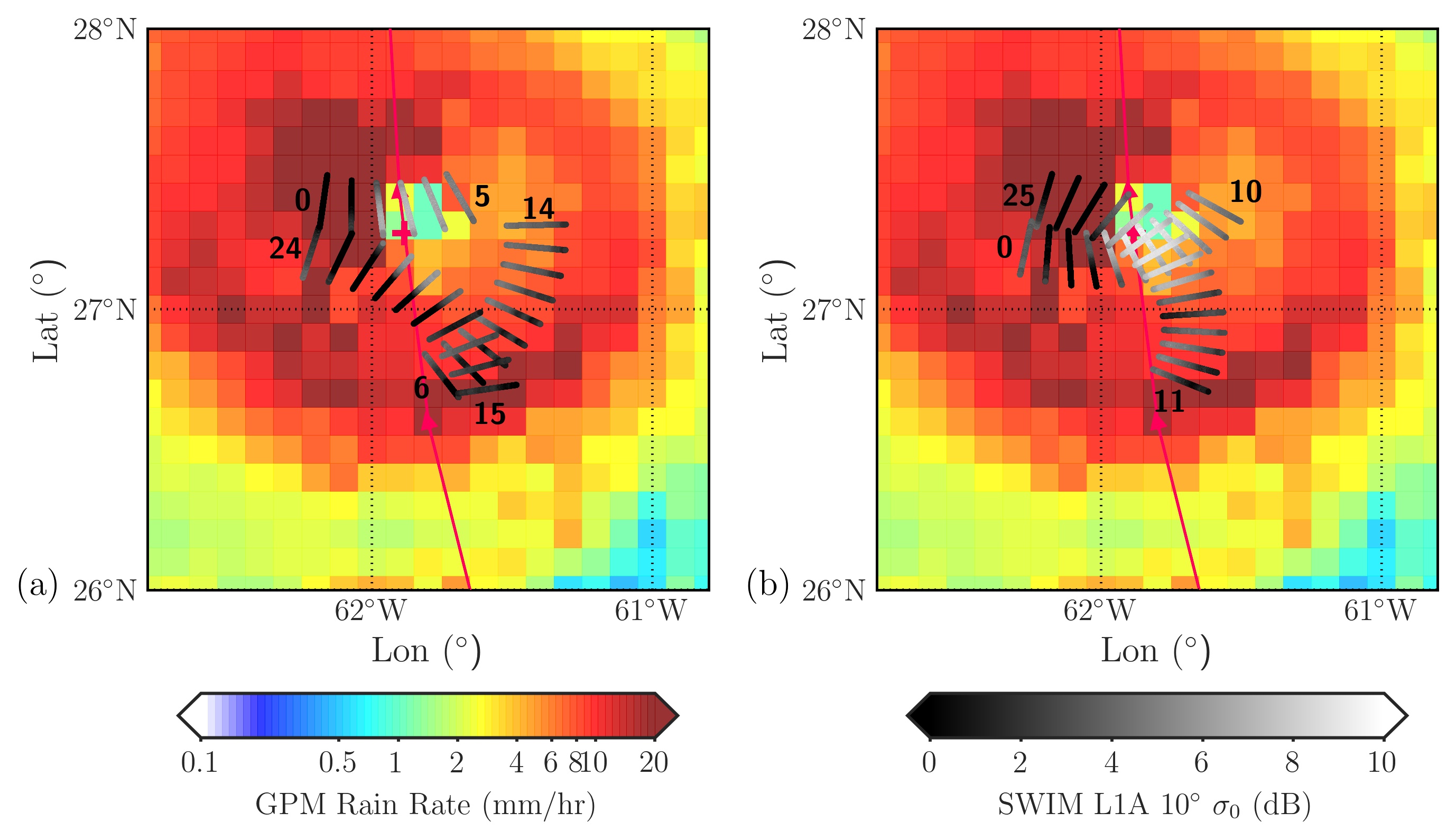}
 \caption{Footprints corresponding to the (a) $10^\circ$ and (b) $8^\circ$ beams, in the box 78-1, superimposed to the rain rate field (in colors) on 1st October 11:30. Red line: trajectory
of the center of SAM. Red cross: location of the center of SAM during the passage of SWIM.}
 \label{fig_swaths_8and10}
\end{figure}

The investigation of the rain impact on the rest of the modulation spectrum (i.e. for $k>k_c$), including the waves energy, deserves a slightly different methodology, relying on both $8^\circ$ and $10^\circ$ spectral beams. Indeed, since this part of the modulation spectrum contains the waves energy, the rain impact can be assessed only by comparing measurements of the same waves but in different rain conditions. The idea is to compare modulation spectra derived from $8^\circ$ and $10^\circ$ beams corresponding to very close azimuth directions, thus measuring waves propagating in the same direction. Assuming that the distance between both footprints is smaller than the spatial variability of waves, but is larger than the spatial variability of the rain field, the discrepancy between both measurements can be assumed to be due to the rain impact. As an example, the footprints of the $10^\circ$ beam (resp. $8^\circ$ beam) corresponding to the box 78-1 are superimposed to the rain rate field in Figure \ref{fig_swaths_8and10}a (resp. \ref{fig_swaths_8and10}b): whereas they are $22$~km apart, the footprint number $3$ of beam $10^\circ$ in the azimuth direction $166^\circ$ (Figure \ref{fig_swaths_8and10}a) undergoes less rain ($RR \approx 2$~mm/hr) than the footprint number $4$ of beam $8^\circ$ in the very close azimuth direction $163^\circ$ ($RR>7$~mm/hr).

\begin{figure}
 \includegraphics[width=1\textwidth]{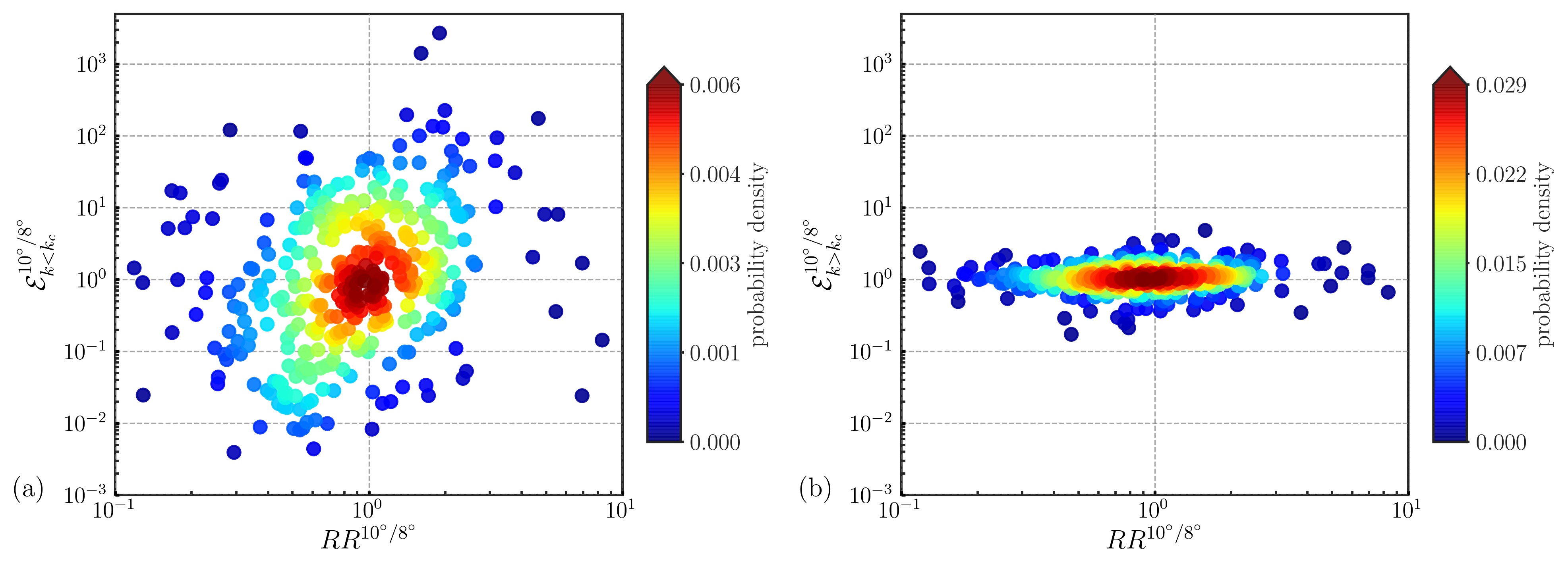}
 \caption{Scatterplot and probability density function (colors) of (a) ${\mathcal E}_{k<k_c}^{10/8^\circ}$ and (b) ${\mathcal E}_{k>k_c}^{10/8^\circ}$  as a function of $RR^{10/8^\circ}$, for $391$ comparisons.}
\label{Ratios_10_8}
\end{figure}

A comparison of the modulation spectra as measured within footprints corresponding to $10^\circ$ and $8^\circ$ beams in very close azimuth directions was undertaken, related to the rain rates. Among the $391$ comparisons, the distance between both footprints varies from $20$~km to $53$~km and the azimuth direction difference from $1.15^\circ$ to $4.30^\circ$. Note that this methodology implies an estimation of rain at the middle of both footprints: hence, the limitations due to the spatial and temporal resolutions of the rain product are now twice as critical compared to Figure \ref{lsvar_Pmsmallk_rain}.  Figure \ref{Ratios_10_8}a shows the ratio of ${\mathcal E}_{k<k_c}$ as measured at $10^\circ$ to that at $8^\circ$ (such a ratio is denoted as ${\mathcal E}_{k<k_c}^{10/8^\circ}$), as a function of the ratio of rain rates, denoted as $RR^{10/8^\circ}$. Despite scattered data, this figure exhibits a correlation between both quantities, which is consistent with Figure \ref{lsvar_Pmsmallk_rain}b. Such a correlation is not observed between ${\mathcal E}_{k>k_c}^{10/8^\circ}$ and $RR^{10/8^\circ}$, as shown in Figure \ref{Ratios_10_8}b. This suggests that rain mainly impacts the modulation spectra at small wavenumbers ($k<k_c$), rather than the waves range. 

\clearpage

\section{KGC15 model for trapped waves}
\label{appendix_KGC15}

The KGC15 model aims at predicting the waves properties of the strongest waves at a given location in the right and left quadrants of a TC using the wind speed $u_w$ and the translation speed of the TC, $V$. The wind speed is supposed to be parallel to the TC axis and uniform along this direction and waves are supposed to propagate along the same direction.  In a stationary TC, the inverse wave age $\alpha_0$ (defined as $u_w/C_p$, where $C_p$ is the phase velocity of the wave) and the dimensionless energy of waves $\tilde{\E}_0$ (using ${u_w^4}/{g^2}$ as unit of energy) are estimated as
\begin{equation} 
\alpha_0=c_\alpha \tilde{\X}^{q}   \qquad {\rm and} \qquad {\tilde{\E}}_0= c_e \tilde{\X}^{p} \,,
\label{Eq_VO}
 \end{equation}
where $\tilde{\X}$ is the dimensionless fetch length (using $u_w^2/g$ as unit of length), measured as the distance from the $y=-ax$ reference line ($\tilde{\X}$ is defined as a positive quantity in both quadrants). In \cite{Kudry15}, $a=1$ (that corresponds to a $45^\circ$ angle); a slightly larger value is chosen here ($a=1.37$, corresponding to a $54^\circ$ angle), in order to extend slightly the range of applicability of the model and thus get more comparisons with observational data. Following \cite{Kudry15}, the fetch law exponents $p$ and $q$ are chosen as $p=0.89$ and $q=-0.275$, which yields $c_\alpha=15.14$ and $c_e=4.41 \times 10^{-7}$. According to (\ref{Eq_VO}), a symmetrical wind field yields a symmetrical wave field. The symmetry is broken in a moving TC, where the inverse wave age obeys
 \begin{equation} 
\left\{
    \begin{array}{ll}
        \alpha^{1/q} \left[1 - \left(1+q\right)^{-1} {\alpha}/{\alpha_T}\right] =  c_\alpha^{1/q} \left[ \tilde{\X}-\tilde{\LL}_{cr}\right] \qquad & \mbox{if} \qquad x>0 \,\, \mbox{and} \,\, \tilde{\ell}>\tilde{\LL}_{cr} \,,\\
         \alpha^{1/q} \left[1 - \left(1+q\right)^{-1} {\alpha}/{\alpha_T}\right] =  c_\alpha^{1/q} \left[ \tilde{\X}-\tilde{\ell}\right] \qquad & \mbox{if} \qquad x>0 \,\, \mbox{and} \,\, \tilde{\ell}<\tilde{\LL}_{cr} \,, \\
         \alpha^{1/q} \left[1 + \left(1+q\right)^{-1} {\alpha}/{\alpha_T}\right]  = c_\alpha^{1/q}\tilde{\X} \qquad & \mbox{if} \qquad x<0 \,,
    \end{array}
\right.
\label{alpha}
\end{equation}
 where  $\tilde{l}$ is the dimensionless length of the segment between the $y=ax$ and $y=-ax$ lines and $\tilde{\LL}_{cr}$ is the dimensionless critical fetch defined as
 \begin{equation} 
 \tilde{\LL}_{cr} = - c_\alpha^{-1/q}\frac{q}{1+q} \alpha_T^{1/q} \,,
 \label{Lcr}
 \end{equation}
 $\alpha_T=u_w/\left(2 V\right)$ being the inverse wave age when their group velocity equals the TC propagation speed $V$. The energy $\tilde{\E}$ is deduced from (\ref{alpha}) through
 \begin{equation} 
   \tilde{\E} = c_e \left({\alpha}/{c_\alpha}\right)^{p/q} \,.
   \label{eqE}
 \end{equation}
 If $\tilde{l}>\tilde{\LL}_{cr}$, waves generated in the right quadrants first travel backward in the frame of the TC, their group velocity increasing until reaching $V$ at a so-called turning point ($\alpha=\alpha_T$), after which they start to propagate forward increasing their group velocity ($\alpha <\alpha_T$) until leaving the TC. These waves have an extended fetch $\tilde{\X}+\tilde{\LL}_{cr}$. If $\tilde{l}<\tilde{\LL}_{cr}$, the waves never reach a group velocity matching $V$ and propagate backward. Finally, in the left quadrants, where the TC propagation is opposite to the TC winds, the waves propagate backward and have a reduced fetch compared to the stationary case. 
 In the model, the TC forcing conditions (wind speed and propagation
 velocity) are supposed to be constant while the waves are propagating in
 the TC.
 The above relations are thus relevant if the TC is quasi-stationary during the time-period $\TT$, 
 this yields an upper bound on the fetch duration
 \begin{equation} 
 \TT \leq \frac{u_w}{g} \frac{2 c_\alpha}{1+q} \left(2\, \tilde{\ell} \right)^{1+q}\,.
 \label{eqT}
 \end{equation}
 In the case of hurricane SAM, at a distance corresponding to
 $R_{max}=28$~km and using $u_w=40$~m/s, (\ref{eqT}) the model thus assumes
 steady winds and a constant translation velocity over a time $\TT \leq 6.8$~hours.

   \bibliographystyle{apa-good.bst}
   \bibliography{biblio}

\end{document}